\documentclass[aps,prl,twocolumn,letterpaper,superscriptaddress,10pt]{revtex4-2}
\usepackage{amssymb,amsthm,amsmath,amsfonts,physics,algorithm2e}
\usepackage{graphicx,ulem,soul,enumerate}
\usepackage[pdftex,dvipsnames,usenames]{xcolor}
\usepackage[colorlinks=true,urlcolor=blue,citecolor=blue,linkcolor=blue]{hyperref}

\begin{document}

\title{Experimental multi-state quantum discrimination through a Quantum network}

\author{Alessandro Laneve}
\email{alessandro.laneve@uniroma1.it}
\affiliation{Dipartimento di Fisica, Sapienza Universit\`{a} di Roma, Piazzale Aldo Moro, 5, I-00185 Roma, Italy}

\author{Andrea Geraldi}
%\email{andrea.geraldi@uniroma1.it}
\affiliation{Dipartimento di Fisica, Sapienza Universit\`{a} di Roma, Piazzale Aldo Moro, 5, I-00185 Roma, Italy}

\author{Frenkli Hamiti}
%\email{hamiti.1639418@studenti.uniroma1.it}
\affiliation{Dipartimento di Fisica, Sapienza Universit\`{a} di Roma, Piazzale Aldo Moro, 5, I-00185 Roma, Italy}

\author{Paolo Mataloni}
%\email{paolo.mataloni@uniroma1.it}
\affiliation{Dipartimento di Fisica, Sapienza Universit\`{a} di Roma, Piazzale Aldo Moro, 5, I-00185 Roma, Italy}

\author{Filippo Caruso}
\email{filippo.caruso@unifi.it}
\affiliation{Department of Physics and Astronomy \& LENS, University of Florence, Via Carrara 1, I-50019 Sesto Fiorentino, Italy}

\date{\today}

\begin{abstract}
    The need of discriminating between different quantum states is a fundamental issue in Quantum Information and Communication. The actual realization of generally optimal strategies in this task is often limited by the need of supplemental resources and very complex receivers.
    We have experimentally implemented two discrimination schemes in a minimum-error scenario based on a receiver featured by a network structure and a dynamical processing of information. The first protocol implemented in our experiment, directly inspired to a recent theoretical proposal, achieves binary optimal discrimination, while the second one provides a novel approach to multi-state quantum discrimination, relying on the dynamical features of the network-like receiver. This strategy exploits the arrival time degree of freedom as an encoding variable, achieving optimal results, without the need for supplemental systems or devices. Our results further reveal the potential of dynamical approaches to Quantum State Discrimination tasks, providing a possible starting point for efficient alternatives to current experimental strategies. 
\end{abstract}

\maketitle

\textit{Introduction ---}
%General introduction:
%\begin{itemize}
%    \item Quantum state discrimination %problem, various methods and %applications.
%    \item Exploitation of quantum neural %networks for quantum tasks.
%\end{itemize}
%Specific protocol:
%\begin{itemize}
%    \item implementation
%    \item application to 4 state %discrimination
%    \item proof of principle + perspective
%\end{itemize}
Quantum State Discrimination (QSD) consists of the ability of distinguishing between the possible states of a quantum
system.
This task can not be achieved in a perfect way in the case of non-orthogonal states.
As a consequence, the quest for general strategies to achieve optimal QSD has become a main issue in Quantum Information topics, regarding its tight bond with the implementation of secure Quantum Communication protocols \cite{chefles2000quantum,bae2015quantum}, or its implications in Quantum Foundations \cite{pusey2012reality, schmid2018contextual}. 
Different approaches can be adopted in QSD: to discriminate states with the minimum error probability \cite{helstrom1976quantum}, or in a non ambiguous way \cite{ivanovic1987differentiate} but admitting unconclusive results.
Other approaches are also possible \cite{croke2006maximum, holevo1973bounds, slussarenko2017quantum}.\\
The quest for effective protocols has led to many results concerning optimal theoretical bounds and optimal receiver models for binary discrimination in different contexts \cite{holevo1974remarks,yuen1975optimum,helstrom1976quantum, acin2005multiple}.
On the other hand, regarding the general case of multiple states and dimension $D>2$, definitive results have not been achieved yet; while the optimal one-shot bound has been set \cite{yuen1975optimum}, many attempts have been done to develop new and more effective strategies %\textcolor{red}{in specific contexts like Quadrature Phase shift Keying (QPSK)  et similia,}
using adaptive protocols \cite{becerra2013experimental,becerra2013implementation,ferdinand2017multi,burenkov2020time} or auxiliary systems \cite{sidhu2021quantum}.
%\textcolor{red}{Out of QPSK context,} 
Other proposals involve the exploitation of an auxiliary system \cite{clarke2001experimental}, in order to increase the dimension of the system to be discriminated, or consist of encoding the states in a complex modal structure \cite{solis2017experimental}. %\textcolor{red}{All of these experimental strategies could be only applied to communication protocols relying on the exploitation of coherent states.}\\
Since adaptive strategies have been proved to be effective, although quite expensive in terms of resources, recent theoretical efforts have been developed to apply neural networks models \cite{patterson2021quantum,caruso2020qsd} and Machine Learning (ML) protocols \cite{fanizza2019optimal} to QSD problems.
In addition to the attempts mentioned above, a dynamical approach based on information processing via Quantum Walks (QWs) has been fancied in the last years \cite{kurzynski2013quantum, li2019implementation}. The network depicted in \cite{caruso2020qsd}, relying on a generalization of QWs, Quantum Stochastic Walks (QSWs) \cite{whitfield2010quantum}, frames a very intuitive model of information processing as well as a wide applicability.\\
In the present work, we experimentally implement a discrimination protocol tightly related to the one proposed in \cite{caruso2020qsd}. %\textcolor{red}{although featuring a few more constraints and a discrete time dynamics. Nevertheless,} 
We demonstrate the effectiveness of a receiver relying on a QSW-like evolution and we experimentally analyze the dynamics of information in a time-binned extraction protocol.
%\textcolor{red}{The second part of the work deals with}
Besides, we report on the realization of an optimal quantum state discrimination protocol for a 4-state alphabet in $D=2$. %\textcolor{red}{, without the exploitation of supplemental resources, such as auxiliary systems or any expansion of the Hilbert space of the system.}
Thanks to the dynamical features of the network, it is possible to encode the quantum information on the four states in the classical observable of detection time, mapping the four different states in four different time-wise classical probability distributions, enhancing their distinguishability.
%\textcolor{red}{These results pave the way for a wide range of extensions, concerning the increase of complexity of the network and the adoption of Machine Learning (ML) methods for the optimal choice of the network parameters.}\\
%\textcolor{red}{On the other hand,}
The discrimination method proposed here may reveal useful in different QSD frameworks, from unambiguous state discrimination to minimum error discrimination (which is the one adopted in this work), and possibly in hybrid protocols, providing an essentially new approach to the problem, featured by an evident spare in terms of resources.

\begin{figure}[!h]
   \centering
    \includegraphics[scale = 0.5]{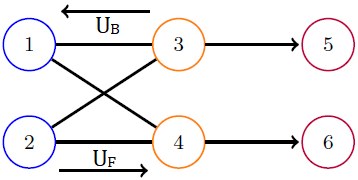}
    \caption{\textbf{2r-2r-2 model of the quantum network.}\textit{ The system goes from a superposition state of input layer nodes $\{1,2\}$ to one of $\{3,4\}$, evolving according to the unitary evolution $\hat{U}_F$. Then, the population of the system which has not been trapped in the sinks evolves through $\hat{U}_B$ to a new general state of the input layer nodes, starting a new evolution loop.}}
    \label{fig:network}
\end{figure}

\textit{Theoretical and experimental framework ---}
%Description of actually realized graph.
%Description of general evolution.
%Description of experimental setup.
%FIGURE: extended experimental setup.
The fundamental aim of our work is to experimentally reproduce a network featured by a 2r-2r-2 topology (according to the nomenclature adopted in \cite{caruso2020qsd}) and depicted in Fig.\ref{fig:network}: two input nodes, namely $\{1,2\}$, are linked in an undirected way with two nodes in the intermediate layer, namely $\{3,4\}$, which are in turn linked in a directed way with two output nodes, namely $\{5,6\}$, acting as sink nodes.
\indent
At variance with the theoretical scheme of \cite{caruso2020qsd}, we propose a model featuring a discrete time evolution and preventing the permanence of the system in the same state after an evolution step.
In conclusion, the state of the network switches from a superposition state of the input nodes to a superposition of the intermediate ones, but for the loss of total probability to the sinks, which eventually brings the network dynamics to vanish.
This model represents a simplification of the original one, but it preserves many of its interesting features: in fact, the dynamical achievement of binary discrimination at the Helstrom bound level remains possible, as well as other implications related to the exploitation of the time dimension.

 \begin{figure}[!h]
        \includegraphics[width=\columnwidth]{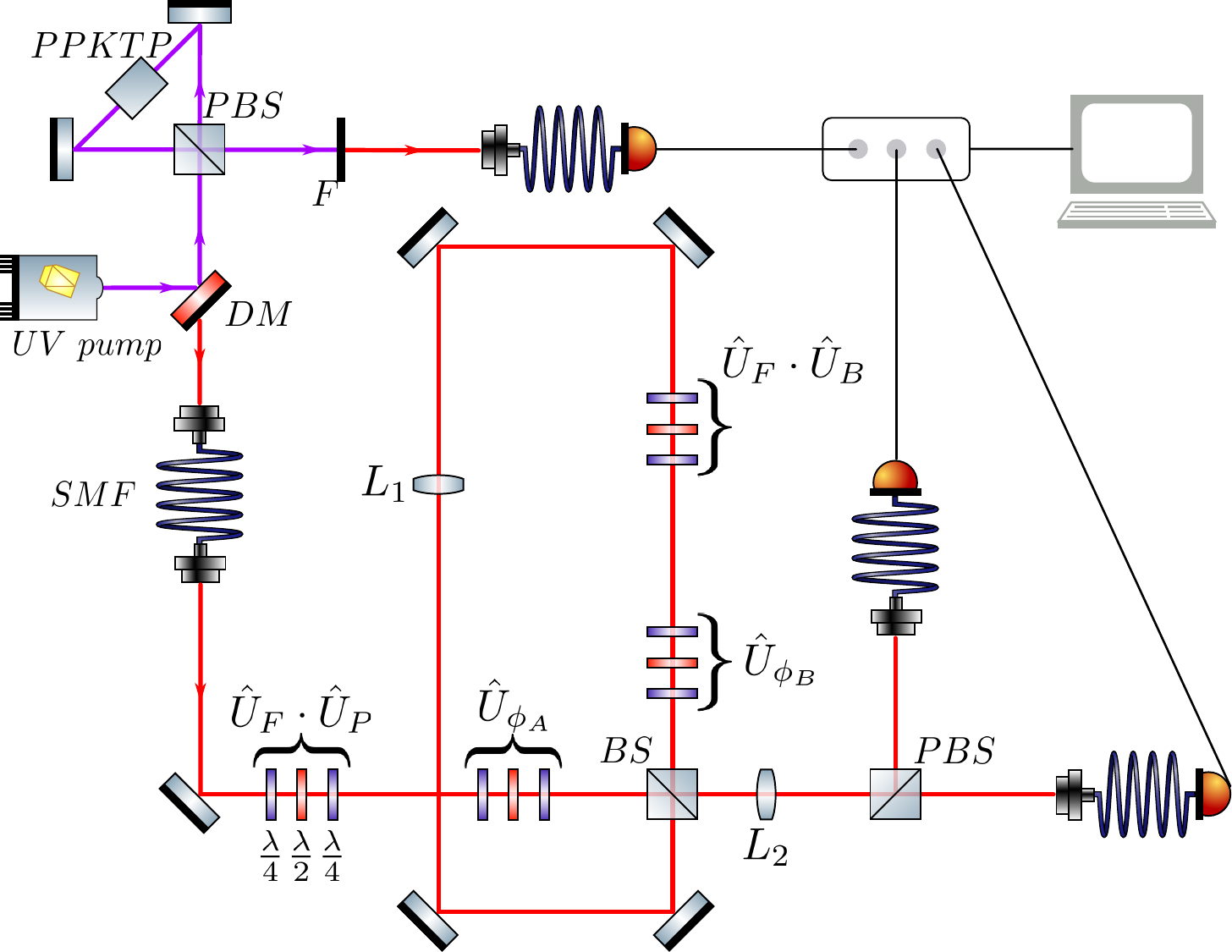}
        \caption{\textbf{Experimental setup realizing the 2r-2r-2 network.} \textit{ The whole unitary operator $\hat{U}=\hat{U}_F\circ \hat{U}_P$ is actually encoded in a single QWP-HWP-QWP set, in order to reduce losses and systematic errors. Each of the optical elements imposes a phase shift $\phi_x$ between the different polarization components. The phase shifts are compensated by the supplemental waveplates sets $\hat{U}_{\phi_{A/B}}$. One lens, identified by $L_1$, is positioned along the loop to prevent losses due to beam divergence; a second one, $L_2$, is located along the extraction path to allow photon collection through multi-mode fibers.
        %Supplemental sets of waveplates L$_{\phi_A}$ and L$_{\phi_B}$ are also placed along the evolution to compensate for the undesired phase difference between $\ket{H}$ and $\ket{V}$ states due to the mirrors along the evolution.
        } }
        \label{fig:setup}
\end{figure}
    
The setup works considering the two input nodes $1,2$ as the two basis states of the polarization of a photon, namely $1 \equiv \ket{H}$ and $2\equiv \ket{V}$, while the photon itself corresponds to the walker of this Quantum Walk-like propagation model.
The initial state of the system $\ket{\psi}$ is encoded in a superposition state of photon polarization.\\
In the experiment, a high brilliance SPDC source $S$ is exploited to generate pairs of photons in a polarization state $\ket{H}_s \otimes \ket{H}_a\equiv \ket{H,H}$, where the subscript $s$ indicates the \textit{system} photon, which will evolve through the network, while the subscript $a$ indicates the \textit{ancilla} photon, which doesn't undergo the network evolution and is exploited as an external trigger for detection.
Details about photon source as well as experimental setup shown in Fig.\ref{fig:setup} are reported in Supplemental Material (SM).\\
Before starting the evolution in the network, the initial state $\ket{\psi}$ is prepared by a unitary preparation stage ($\hat{U}_{P}$).
In our setup, unitary operations are applied through different waveplates sets composed by a sequence of a quarter waveplate (QWP), a half waveplate (HWP) and a further QWP.
This sequence allows to apply any kind of unitary operator in the polarization degree of freedom, allowing to transform the state $\ket{H}_s$ into any desired pure $\ket{\psi}$ state.
After the state preparation, the system can be considered to be in the input layer of the network.
The photon is then subjected to the first evolution stage, applying the operator $\hat{U}_F$ for the evolution from nodes $\{1,2\}$ to $\{3,4\}$.
The nodes $\{3,4$\}, also referred to as \textit{sinker nodes}, are identified with the basis states of the polarization, too.
The time degree of freedom allows to discriminate between states in the nodes $\{1,2$\} and the ones in $\{3,4$\}.
A beam splitter (BS) is then placed along the path: its aim is to redirect the photon to the sink nodes $\{5,6\}$ with a given probability $p_s$ or to send it into the network with probability $1-p_s$, in order to continue the evolution.
In the first case, when the photon travels towards the sinks, it impinges on a Polarizing Beam Splitter (PBS) which separates the $\ket{H}$ and $\ket{V}$ components of the photon state, representing the sink nodes $\{5,6\}$.
The population of both sinks is then measured by single photon detectors.
The detection of the $s$ photon is performed in coincidence with the detection of the $a$ photon, thanks to a preliminary synchronization process.
In the second case, i.e. if the photon is redirected into the network, it passes through another set of QWP-HWP-QWP, which are set to apply the composite operator $\hat{U}_F\circ\hat{U}_B$, where $\hat{U}_B$ is the operator describing the propagation of the system from nodes $\{3,4\}$ to $\{1,2\}$.
In this way the system evolves from a state of nodes $\{3,4\}$ to $\{1,2\}$ and then again to nodes $\{3,4\}$.
%Since the walker can not travel from nodes $\{1,2\}$ to sinks $\{5,6\}$ without passing again through nodes $\{3,4\}$, the waveplates set L$_S$  $\{1,2\}$ to $\{3,4\}$.
After that, the walker impinges again on the BS, beginning another loop evolution or being detected into the sinks.
The above setup provides an experimental realization of a network featured by a 2r-2r-2 topology.

%A removable tomography setting (RTS) is also placed at need right to be sure that the initial state is correctly set.\\
%A main drawback of the setup consists of the following: at the first interaction of the BS, the probability for the photon to go to the sinks is $p_s=T$, where $T$ is the transmission coefficient of the BS, while the probability to continue the evolution is $1-p_s = 1- T = R$, where $R$ is the reflection coefficient of the $BS$.
%For all subsequent interactions, the situation is the opposite: the probability to go to sinks becomes $R$ while the one to continue the evolution is $T$.

\begin{figure}[!t]
        \includegraphics[width=\columnwidth]{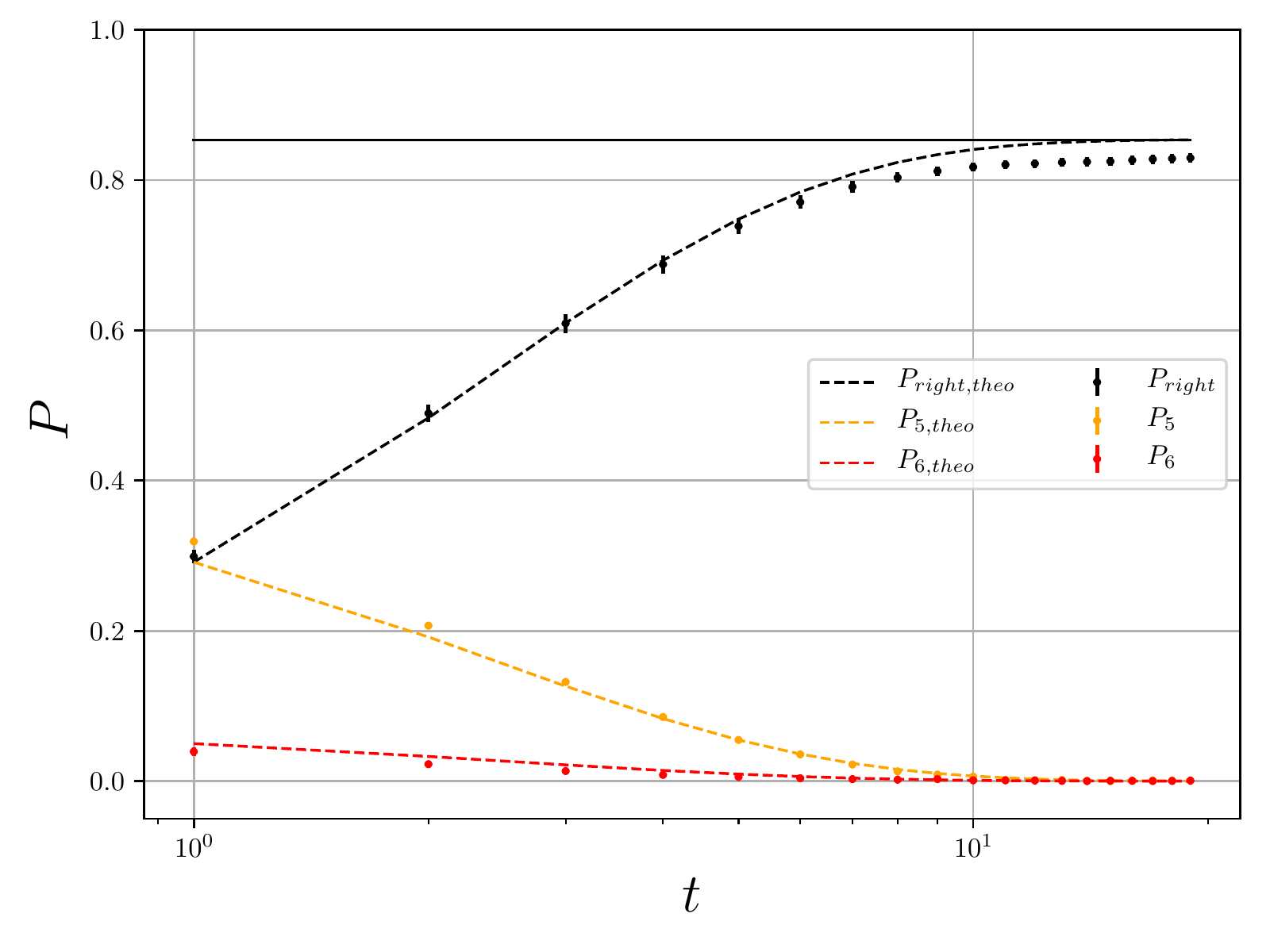}
        \caption{\textbf{Cumulative curve and output probability in time} \textit{Temporal evolution of the experimental output probability $P_5(6)$ in sink $5(6)$ for the input state $\ket{\psi1}$, together with the experimental average cumulative curve $P_{right}$, corresponding to the total probability of correct discrimination. The experimental results are reported in comparison with the numerical expectations.
        Black solid line represents the Helstrom Bound computed for the two selected states. A systematic error in the state preparation leads to an imperfect match of the bound, as the extracted information saturates. 
        } }
        \label{fig:binary_counts}
\end{figure}

\textit{Experimental realization of binary quantum state discrimination---}
%Description of discrimination protocol %(cumulative).
%Numerical and experimental results.
%FIGURE: cumulative curve num and exp.
Through the experimental apparatus described above it is possible to implement a discrimination protocol quite similar to the one proposed in \cite{caruso2020qsd}.
Two non-orthogonal states $\{\ket{\psi_1},\ket{\psi_2\}}$ are encoded in two states of the input layer; in order to discriminate them, the unitary evolution of the network must be tailored in such a way that the detection of the system in a sink reveals with the minimum uncertainty the presence of a given input state rather than the other one, with the minimum uncertainty.
Such an optimal network will be able to discriminate states up to the Helstrom bound, by the computation of the cumulative population of the sinks in time.
The states experimentally exploited for the protocol are $\ket{\psi_1}=\cos(\frac{\pi}{8})\ket{H}+\sin(\frac{\pi}{8})\ket{V}$ and $\ket{\psi_2}=\cos(\frac{\pi}{8})\ket{H}-\sin(\frac{\pi}{8})\ket{V}$, setting a strong similarity with the case studied in \cite{caruso2020qsd}.
The optimal unitary matrix for the realization of a suitable dynamical discrimination protocol was analytically found by maximizing or minimizing the output probability of each sink with respect to the corresponding input state. 
The result of this optimization lead to the evolution matrix $U=U_F=U_B=\frac{1}{\sqrt{2}}\begin{pmatrix}
1 & 1\\
1 & -1
\end{pmatrix}$, which has some analogy to the one computed in \cite{caruso2020qsd}, besides some peculiar property. In fact, exploiting this evolution matrix, the discrimination protocol consists of a repeated optimal single-shot discrimination protocol: the network periodically brings the system in the state of the sinker nodes allowing optimal discrimination.
In this way, at each loop completion, the information on the state is optimally extracted, leading to the output probabilities in time $P_5$ and $P_6$ reported in Fig.\ref{fig:binary_counts}, where the sink node 5 is associated to the detection of state $\ket{\psi_1}$, while sink 6 to the detection of $\ket{\psi_2}$.
It is now interesting to understand the dynamical behaviour of the probability of correct guess (as well as the dynamics of the extracted information).
To this aim, the cumulative probability of correct discrimination in time is computed, for both states: $C_1=\sum_\tau^tp_5^{(1)}(\tau)$, $C_2=\sum_\tau^tp_6^{(2)}(\tau)$, where $p_k^{(i)}(\sigma)$ is the population of sink $k$, with input state $\ket{\psi_i}$, measured after the $\sigma$th loop.

The curve resulting from the average between $C_1$ and $C_2$, normalized to the total cumulative probability, is exhibited in Fig.\ref{fig:binary_counts} as $P_{right}$, in comparison with the numerical expectations. 
The experimental analysis proves the effectiveness of this protocol to achieve Helstrom level binary quantum state discrimination in a dynamical context.
This proof paves the way towards the exploitation of these dynamical features in more complex and powerful protocols, fully taking advantage of a time-binned extraction of information.
\\

\textit{Experimental multi-state discrimination via time-binning protocol---}
%Description of the problem of 4 state %discrimination.
%Theoretical description of discrimination %protocol and how "not really optimal" %unitaries are found, both for Compass and %Tetrad.
%Numerical and experimental results.
%FIGURE: time signatures num and exp, %compass and tetrad.
%Highlights on applications, originality and %actual gain. Study in function of exploited %copies of the state.
%FIGURE: scaling num and exp, compass and %tetrad (the best one, as an instance).
%\\
The discrimination of more than two non-orthogonal states in a 2-dimensional space is, in general, a task for which an absolutely optimal strategy is not known \cite{bae2015quantum}.
We have developed a protocol based on time-binning that can be exploited to discriminate different sets of four states without accessing auxiliary systems or resources.
 \begin{figure}[!h]
        \includegraphics[width=\columnwidth]{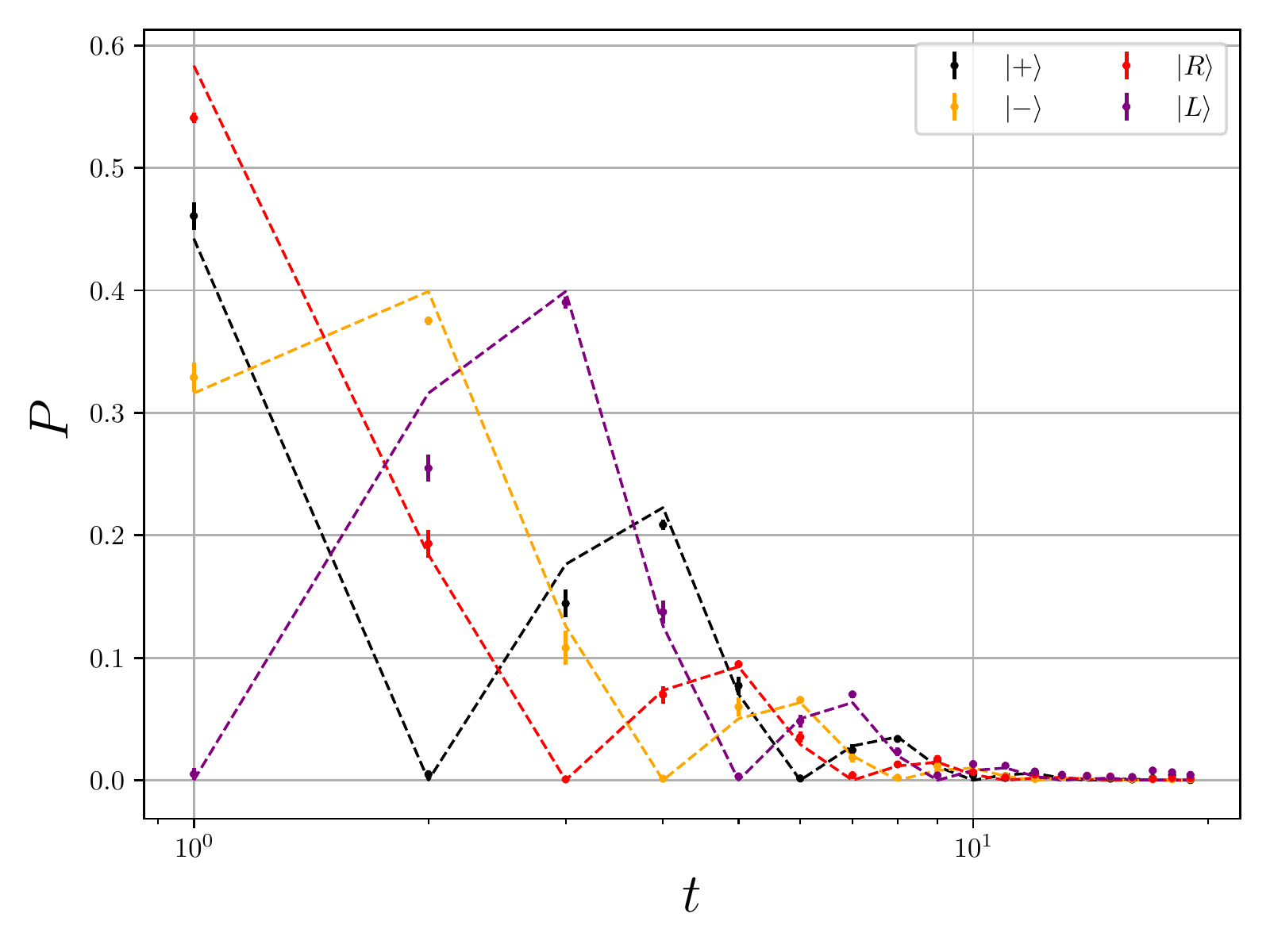}\\
        \includegraphics[width=\columnwidth]{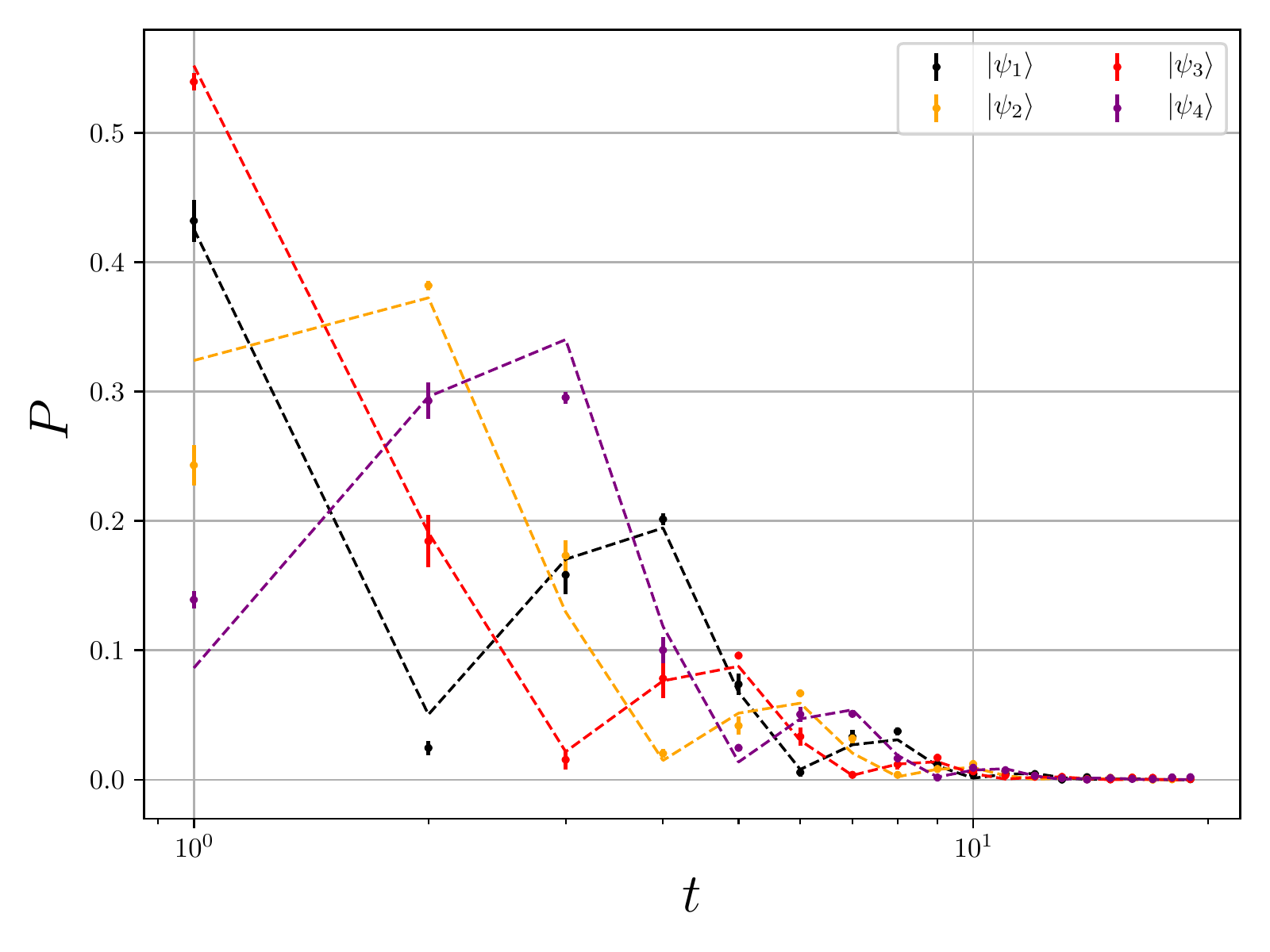}
        \caption{\textbf{Step-wise output probability for geometrically uniform states and the Tetrad set.} \textit{Experimental output probability as a function of the number of round trips travelled by photons when the geometrically uniform states (top) and the Tetrad (bottom) states circulate in the network. Normalization of each distribution is performed summing over the total output probability for each single state, in order to account for the experimental signal decrease as the observed time bin grows. Data are reported in comparison with corresponding numerical results (dashed lines).
        } }
        \label{fig:timesig_compass}
\end{figure}
\\
\indent
The protocol consists of a redistribution of probability in time by means of the dynamical features of the network.
The unitary evolution operators $\hat{U}_F$ and $\hat{U}_B$ are chosen such that each of the four states is featured by a detection probability distribution in time maximally different each from the others.
The first observed application involves a set of geometrically uniform states \cite{bae2015quantum}, which we consider equally likely to be received:
\begin{equation}\label{eq:compass}\begin{split}
&\ket{+}=1/\sqrt{2}\,(\ket{H}+\ket{V})\\
&\ket{-}=1/\sqrt{2}\,(\ket{H}-\ket{V})\\
&\ket{R}=1/\sqrt{2}\,(\ket{H}+i\ket{V})\\
&\ket{L}=1/\sqrt{2}\,(\ket{H}-i\ket{V})\\
\end{split}\end{equation} 
The suitable unitary evolution matrices were identified as $U_F=\frac{1}{\sqrt{2}}\begin{pmatrix}
1 & 1\\
1 & -1
\end{pmatrix}$ and $U_B=\frac{1}{\sqrt{2}}\begin{pmatrix}
\frac{1+i}{\sqrt{2}} & \frac{1+i}{\sqrt{2}}\\
\frac{1-i}{\sqrt{2}} & \frac{i-1}{\sqrt{2}}
\end{pmatrix}$, through an analytical method described in SM.
The resulting experimental detection probability distributions in time are reported in Fig. \ref{fig:timesig_compass} top, in comparison with the numerical ones.

The main feature of these distributions, each of which provides an unambiguous time-signature of a given input state, is the 4-periodicity of the probability trajectory in time; it relies on the network structure of our receiver and is intimately connected to the number of intermediate layers of it.
A more detailed discussion about the actual implemented projectors and their bond with the network structure is reported in Supplemental Material (SM).
This particular case of study features two pairs of orthogonal states, providing an intrinsic advantage in producing maximally different time dependent distributions.
The same protocol was also applied to the set known as Tetrad set \cite{clarke2001experimental}, consisting of four maximally equidistant states on the Poincaré sphere:
\begin{equation}\label{eq:tetrad}\begin{split}
&\ket{\psi_1}=1/\sqrt{3}\,(-\ket{H}+\sqrt{2}e^{-2\pi i/3}\ket{V}) \\
&\ket{\psi_2}=1/\sqrt{3}\,(-\ket{H}+\sqrt{2}e^{2\pi i/3}\ket{V})\\
&\ket{\psi_3}=1/\sqrt{3}\,(-\ket{H}+\sqrt{2}\ket{V})\\
&\ket{\psi_4}=\ket{H}\\
 \end{split}\end{equation}
featuring several interesting properties for quantum communication and cryptography.
The corresponding results are given in Fig. \ref{fig:timesig_compass} bottom.
In this case, the protocol is slightly suboptimal, since the distribution minima are not vanishing, but a deeper optimization procedure may return more effective results.
The time dependent distributions show a good agreement with the numerical expectations computed accounting for actual parameters, demonstrating a reliable procedure of multi-state discrimination,  without the exploitation of any supplementary system, or spreading the states in multi-mode configurations which require an abundance of detecting devices.
Indeed, by this procedure, a single photon-counting detector is needed (the second one only provides complementary information), together with the capability of discriminating the arrival time of photons with quite loosen precision.
In realistic communication protocols, the second trigger photon can be replaced by more sophisticated synchronization techniques.
An experimental analysis of the efficiency of the protocol described by these time distributions in a multi-copy framework is reported in SM.%: for the geometrically homogeneous set, an average one-shot measure achieves the optimal bound stated in \cite{yuen1975optimum}, while the result is slightly suboptimal for the Tetrad set.
%Besides, it is also possible to have a glance at the scaling of the error probability as the average number of photons increases (more details are provided in SM).
\\

\textit{Discussion and conclusion---}
%State of the art for single photons state %discrimination.
%Exploitation of a network.
%The needed projectors overlap (in time %dimension), generating a resources spare.
%Perspectives:
%\begin{itemize}
%    \item exploitation of ML for optimizing %the protocol in more general cases and %for discrimination among families of %states
%    \item extension to an higher number of %states via deeper networks
%    \item exploitation of actual %single-photon sources
%\end{itemize}
%Pointing out again the completely new %approach, which doesn't need ancillary %systems or the distribution of states over %supplementary modes.
In the last two decades genunine quantum features have been often exploited as a catalyst to improve the efficiency of a number of practical tasks. In this scenario, optimal strategies for Quantum State Discrimination of actual quantum states have a great deal of relevance and a wide range of applications: Quantum Communication  \cite{fujiwara2003exceeding}, Quantum Key Distribution \cite{cerf2002security} and also Quantum Sensing, in the case of distinguishing different external fields affecting the system dynamics (such as in NV-center noise spectroscopy \cite{degen2017} or avian magnetoreception \cite{lambert2013quantum}).
In recent years the development of actual single-photon protocols for QSD has been quite rare, while a lot of effort has been spent in developing protocols requiring the exploitation of coherent states (adaptive strategies, Quantum Phase Shift Keying \cite{becerra2013experimental,becerra2013implementation,ferdinand2017multi,sidhu2021quantum}). However, it is well known that the use of coherent states in Quantum Communication protocols does not grant equal security as compared to the exploitation of actual quantum states. \\
In this work, we exploited actual single-photon states and we achieved a nearly optimal protocol featuring a clear spare of resources, in terms of auxiliary systems and physical measurement devices.
The spare of auxiliary physical systems is achieved thanks to the network structure of our receiver, which allows us to implement strategies based on time-binning of information extraction, in the case of binary and multi-state discrimination, without the need for supplemental devices; that is a relevant property when regarding for applications. 
The exploitation of actual quantum states in this protocol makes it quite interesting for the application to secure Quantum Communication tasks and Quantum Key Distribution, tightly relying on the quantum nature of the implementing platforms for their effective realization.
Our results represent on one side a basic proof of the protocols effectiveness, but they pave the way to further extensions, such as the adoption of adaptive methods for more general tasks, or the implementation of more complex networks, in order to increase the maximum possible number of states which could be discriminated.\\
Through these improvements, we will be able to turn our simple model in a proper Quantum Neural Network \cite{schuld2014quest}, with a potential which is yet to be uncovered in this field.
Noise-robustness of our protocols needs still to be studied, consistently with the approach of \cite{caruso2020qsd}, in order to reveal possible usefulness for Quantum Computing in NISQ devices. Even Machine Learning techniques need to be applied to our framework, aiming at the development of quantum machine learning protocols in which quantum information is classified as in classical supervised deep learning schemes.
\\
\indent In conclusion, a new approach in experimental realizations of QSD protocols has been reported in this work, featuring some practical and theoretical advantages for the applications and leaving a lot of room for improvement and generalization. It may represent a valuable alternative to established strategies, once its potential is fully understood.
\textit{Acknowledgments.---} We acknowledge support from MIUR (Ministero dell’Istruzione, dell’Università e della Ricerca) via project PRIN 2017 “Taming complexity via QUantum Strategies a Hybrid Integrated Photonic approach” (QUSHIP) 
Id. 2017SRNBRK. F.C. was financially supported by the PATHOS EU H2020 FET-OPEN Grant No. 828946, the Fondazione CR Firenze through the project QUANTUM-AI, and the Florence University Grant Q-CODYCES.

%European Commission Grants No. FP7-ICT-2011-9-600838 (QWAD–Quantum Waveguides Application and Development).

%Ideas to make the conclusions more interesting for %a broader audience:
%
%-Quantum Communication
%
%-QKD
%
%-generalization to quantum neural networks
%
%-quantum state discrimination for quantum %computing
%
%-noise-robustness shown in [21] useful for NISQ %devices
%
%-it can be applied to quantum sensing where QSD is %exploited to distinguish different external (e.g., %magnetic) fields affecting the system dynamics -- %see NV-center noise spectroscopy or even what %Nature performs via the avian magnetic compass in %bird navigation. 
%
%-ML algorithms to optimize the discrimination in a %forthcoming work
%
%-First steps towards quantum machine learning %protocols where quantum information is classified %as in classical supervised deep learning schemes.

\bibliography{biblio}

\begin{thebibliography}{30}%
\makeatletter
\providecommand \@ifxundefined [1]{%
 \@ifx{#1\undefined}
}%
\providecommand \@ifnum [1]{%
 \ifnum #1\expandafter \@firstoftwo
 \else \expandafter \@secondoftwo
 \fi
}%
\providecommand \@ifx [1]{%
 \ifx #1\expandafter \@firstoftwo
 \else \expandafter \@secondoftwo
 \fi
}%
\providecommand \natexlab [1]{#1}%
\providecommand \enquote  [1]{``#1''}%
\providecommand \bibnamefont  [1]{#1}%
\providecommand \bibfnamefont [1]{#1}%
\providecommand \citenamefont [1]{#1}%
\providecommand \href@noop [0]{\@secondoftwo}%
\providecommand \href [0]{\begingroup \@sanitize@url \@href}%
\providecommand \@href[1]{\@@startlink{#1}\@@href}%
\providecommand \@@href[1]{\endgroup#1\@@endlink}%
\providecommand \@sanitize@url [0]{\catcode `\\12\catcode `\$12\catcode
  `\&12\catcode `\#12\catcode `\^12\catcode `\_12\catcode `\%12\relax}%
\providecommand \@@startlink[1]{}%
\providecommand \@@endlink[0]{}%
\providecommand \url  [0]{\begingroup\@sanitize@url \@url }%
\providecommand \@url [1]{\endgroup\@href {#1}{\urlprefix }}%
\providecommand \urlprefix  [0]{URL }%
\providecommand \Eprint [0]{\href }%
\providecommand \doibase [0]{https://doi.org/}%
\providecommand \selectlanguage [0]{\@gobble}%
\providecommand \bibinfo  [0]{\@secondoftwo}%
\providecommand \bibfield  [0]{\@secondoftwo}%
\providecommand \translation [1]{[#1]}%
\providecommand \BibitemOpen [0]{}%
\providecommand \bibitemStop [0]{}%
\providecommand \bibitemNoStop [0]{.\EOS\space}%
\providecommand \EOS [0]{\spacefactor3000\relax}%
\providecommand \BibitemShut  [1]{\csname bibitem#1\endcsname}%
\let\auto@bib@innerbib\@empty
%</preamble>
\bibitem [{\citenamefont {Chefles}(2000)}]{chefles2000quantum}%
  \BibitemOpen
  \bibfield  {author} {\bibinfo {author} {\bibfnamefont {A.}~\bibnamefont
  {Chefles}},\ }\bibfield  {title} {\bibinfo {title} {Quantum state
  discrimination},\ }\href@noop {} {\bibfield  {journal} {\bibinfo  {journal}
  {Contemporary Physics}\ }\textbf {\bibinfo {volume} {41}},\ \bibinfo {pages}
  {401} (\bibinfo {year} {2000})}\BibitemShut {NoStop}%
\bibitem [{\citenamefont {Bae}\ and\ \citenamefont
  {Kwek}(2015)}]{bae2015quantum}%
  \BibitemOpen
  \bibfield  {author} {\bibinfo {author} {\bibfnamefont {J.}~\bibnamefont
  {Bae}}\ and\ \bibinfo {author} {\bibfnamefont {L.-C.}\ \bibnamefont {Kwek}},\
  }\bibfield  {title} {\bibinfo {title} {Quantum state discrimination and its
  applications},\ }\href@noop {} {\bibfield  {journal} {\bibinfo  {journal}
  {Journal of Physics A: Mathematical and Theoretical}\ }\textbf {\bibinfo
  {volume} {48}},\ \bibinfo {pages} {083001} (\bibinfo {year}
  {2015})}\BibitemShut {NoStop}%
\bibitem [{\citenamefont {Pusey}\ \emph {et~al.}(2012)\citenamefont {Pusey},
  \citenamefont {Barrett},\ and\ \citenamefont {Rudolph}}]{pusey2012reality}%
  \BibitemOpen
  \bibfield  {author} {\bibinfo {author} {\bibfnamefont {M.~F.}\ \bibnamefont
  {Pusey}}, \bibinfo {author} {\bibfnamefont {J.}~\bibnamefont {Barrett}},\
  and\ \bibinfo {author} {\bibfnamefont {T.}~\bibnamefont {Rudolph}},\
  }\bibfield  {title} {\bibinfo {title} {On the reality of the quantum state},\
  }\href@noop {} {\bibfield  {journal} {\bibinfo  {journal} {Nature Physics}\
  }\textbf {\bibinfo {volume} {8}},\ \bibinfo {pages} {475} (\bibinfo {year}
  {2012})}\BibitemShut {NoStop}%
\bibitem [{\citenamefont {Schmid}\ and\ \citenamefont
  {Spekkens}(2018)}]{schmid2018contextual}%
  \BibitemOpen
  \bibfield  {author} {\bibinfo {author} {\bibfnamefont {D.}~\bibnamefont
  {Schmid}}\ and\ \bibinfo {author} {\bibfnamefont {R.~W.}\ \bibnamefont
  {Spekkens}},\ }\bibfield  {title} {\bibinfo {title} {Contextual advantage for
  state discrimination},\ }\href@noop {} {\bibfield  {journal} {\bibinfo
  {journal} {Physical Review X}\ }\textbf {\bibinfo {volume} {8}},\ \bibinfo
  {pages} {011015} (\bibinfo {year} {2018})}\BibitemShut {NoStop}%
\bibitem [{\citenamefont {Helstrom}\ and\ \citenamefont
  {Helstrom}(1976)}]{helstrom1976quantum}%
  \BibitemOpen
  \bibfield  {author} {\bibinfo {author} {\bibfnamefont {C.~W.}\ \bibnamefont
  {Helstrom}}\ and\ \bibinfo {author} {\bibfnamefont {C.~W.}\ \bibnamefont
  {Helstrom}},\ }\href@noop {} {\emph {\bibinfo {title} {Quantum detection and
  estimation theory}}},\ Vol.~\bibinfo {volume} {84}\ (\bibinfo  {publisher}
  {Academic press New York},\ \bibinfo {year} {1976})\BibitemShut {NoStop}%
\bibitem [{\citenamefont {Ivanovic}(1987)}]{ivanovic1987differentiate}%
  \BibitemOpen
  \bibfield  {author} {\bibinfo {author} {\bibfnamefont {I.~D.}\ \bibnamefont
  {Ivanovic}},\ }\bibfield  {title} {\bibinfo {title} {How to differentiate
  between non-orthogonal states},\ }\href@noop {} {\bibfield  {journal}
  {\bibinfo  {journal} {Physics Letters A}\ }\textbf {\bibinfo {volume}
  {123}},\ \bibinfo {pages} {257} (\bibinfo {year} {1987})}\BibitemShut
  {NoStop}%
\bibitem [{\citenamefont {Croke}\ \emph {et~al.}(2006)\citenamefont {Croke},
  \citenamefont {Andersson}, \citenamefont {Barnett}, \citenamefont {Gilson},\
  and\ \citenamefont {Jeffers}}]{croke2006maximum}%
  \BibitemOpen
  \bibfield  {author} {\bibinfo {author} {\bibfnamefont {S.}~\bibnamefont
  {Croke}}, \bibinfo {author} {\bibfnamefont {E.}~\bibnamefont {Andersson}},
  \bibinfo {author} {\bibfnamefont {S.~M.}\ \bibnamefont {Barnett}}, \bibinfo
  {author} {\bibfnamefont {C.~R.}\ \bibnamefont {Gilson}},\ and\ \bibinfo
  {author} {\bibfnamefont {J.}~\bibnamefont {Jeffers}},\ }\bibfield  {title}
  {\bibinfo {title} {Maximum confidence quantum measurements},\ }\href@noop {}
  {\bibfield  {journal} {\bibinfo  {journal} {Physical review letters}\
  }\textbf {\bibinfo {volume} {96}},\ \bibinfo {pages} {070401} (\bibinfo
  {year} {2006})}\BibitemShut {NoStop}%
\bibitem [{\citenamefont {Holevo}(1973)}]{holevo1973bounds}%
  \BibitemOpen
  \bibfield  {author} {\bibinfo {author} {\bibfnamefont {A.~S.}\ \bibnamefont
  {Holevo}},\ }\bibfield  {title} {\bibinfo {title} {Bounds for the quantity of
  information transmitted by a quantum communication channel},\ }\href@noop {}
  {\bibfield  {journal} {\bibinfo  {journal} {Problemy Peredachi Informatsii}\
  }\textbf {\bibinfo {volume} {9}},\ \bibinfo {pages} {3} (\bibinfo {year}
  {1973})}\BibitemShut {NoStop}%
\bibitem [{\citenamefont {Slussarenko}\ \emph {et~al.}(2017)\citenamefont
  {Slussarenko}, \citenamefont {Weston}, \citenamefont {Li}, \citenamefont
  {Campbell}, \citenamefont {Wiseman},\ and\ \citenamefont
  {Pryde}}]{slussarenko2017quantum}%
  \BibitemOpen
  \bibfield  {author} {\bibinfo {author} {\bibfnamefont {S.}~\bibnamefont
  {Slussarenko}}, \bibinfo {author} {\bibfnamefont {M.~M.}\ \bibnamefont
  {Weston}}, \bibinfo {author} {\bibfnamefont {J.-G.}\ \bibnamefont {Li}},
  \bibinfo {author} {\bibfnamefont {N.}~\bibnamefont {Campbell}}, \bibinfo
  {author} {\bibfnamefont {H.~M.}\ \bibnamefont {Wiseman}},\ and\ \bibinfo
  {author} {\bibfnamefont {G.~J.}\ \bibnamefont {Pryde}},\ }\bibfield  {title}
  {\bibinfo {title} {Quantum state discrimination using the minimum average
  number of copies},\ }\href@noop {} {\bibfield  {journal} {\bibinfo  {journal}
  {Physical review letters}\ }\textbf {\bibinfo {volume} {118}},\ \bibinfo
  {pages} {030502} (\bibinfo {year} {2017})}\BibitemShut {NoStop}%
\bibitem [{\citenamefont {Holevo}(1974)}]{holevo1974remarks}%
  \BibitemOpen
  \bibfield  {author} {\bibinfo {author} {\bibfnamefont {A.~S.}\ \bibnamefont
  {Holevo}},\ }\bibfield  {title} {\bibinfo {title} {Remarks on optimal quantum
  measurements},\ }\href@noop {} {\bibfield  {journal} {\bibinfo  {journal}
  {Problemy Peredachi Informatsii}\ }\textbf {\bibinfo {volume} {10}},\
  \bibinfo {pages} {51} (\bibinfo {year} {1974})}\BibitemShut {NoStop}%
\bibitem [{\citenamefont {Yuen}\ \emph {et~al.}(1975)\citenamefont {Yuen},
  \citenamefont {Kennedy},\ and\ \citenamefont {Lax}}]{yuen1975optimum}%
  \BibitemOpen
  \bibfield  {author} {\bibinfo {author} {\bibfnamefont {H.}~\bibnamefont
  {Yuen}}, \bibinfo {author} {\bibfnamefont {R.}~\bibnamefont {Kennedy}},\ and\
  \bibinfo {author} {\bibfnamefont {M.}~\bibnamefont {Lax}},\ }\bibfield
  {title} {\bibinfo {title} {Optimum testing of multiple hypotheses in quantum
  detection theory},\ }\href@noop {} {\bibfield  {journal} {\bibinfo  {journal}
  {IEEE Transactions on Information Theory}\ }\textbf {\bibinfo {volume}
  {21}},\ \bibinfo {pages} {125} (\bibinfo {year} {1975})}\BibitemShut
  {NoStop}%
\bibitem [{\citenamefont {Ac{\'\i}n}\ \emph {et~al.}(2005)\citenamefont
  {Ac{\'\i}n}, \citenamefont {Bagan}, \citenamefont {Baig}, \citenamefont
  {Masanes},\ and\ \citenamefont {Munoz-Tapia}}]{acin2005multiple}%
  \BibitemOpen
  \bibfield  {author} {\bibinfo {author} {\bibfnamefont {A.}~\bibnamefont
  {Ac{\'\i}n}}, \bibinfo {author} {\bibfnamefont {E.}~\bibnamefont {Bagan}},
  \bibinfo {author} {\bibfnamefont {M.}~\bibnamefont {Baig}}, \bibinfo {author}
  {\bibfnamefont {L.}~\bibnamefont {Masanes}},\ and\ \bibinfo {author}
  {\bibfnamefont {R.}~\bibnamefont {Munoz-Tapia}},\ }\bibfield  {title}
  {\bibinfo {title} {Multiple-copy two-state discrimination with individual
  measurements},\ }\href@noop {} {\bibfield  {journal} {\bibinfo  {journal}
  {Physical Review A}\ }\textbf {\bibinfo {volume} {71}},\ \bibinfo {pages}
  {032338} (\bibinfo {year} {2005})}\BibitemShut {NoStop}%
\bibitem [{\citenamefont {Becerra}\ \emph
  {et~al.}(2013{\natexlab{a}})\citenamefont {Becerra}, \citenamefont {Fan},
  \citenamefont {Baumgartner}, \citenamefont {Goldhar}, \citenamefont
  {Kosloski},\ and\ \citenamefont {Migdall}}]{becerra2013experimental}%
  \BibitemOpen
  \bibfield  {author} {\bibinfo {author} {\bibfnamefont {F.}~\bibnamefont
  {Becerra}}, \bibinfo {author} {\bibfnamefont {J.}~\bibnamefont {Fan}},
  \bibinfo {author} {\bibfnamefont {G.}~\bibnamefont {Baumgartner}}, \bibinfo
  {author} {\bibfnamefont {J.}~\bibnamefont {Goldhar}}, \bibinfo {author}
  {\bibfnamefont {J.}~\bibnamefont {Kosloski}},\ and\ \bibinfo {author}
  {\bibfnamefont {A.}~\bibnamefont {Migdall}},\ }\bibfield  {title} {\bibinfo
  {title} {Experimental demonstration of a receiver beating the standard
  quantum limit for multiple nonorthogonal state discrimination},\ }\href@noop
  {} {\bibfield  {journal} {\bibinfo  {journal} {Nature Photonics}\ }\textbf
  {\bibinfo {volume} {7}},\ \bibinfo {pages} {147} (\bibinfo {year}
  {2013}{\natexlab{a}})}\BibitemShut {NoStop}%
\bibitem [{\citenamefont {Becerra}\ \emph
  {et~al.}(2013{\natexlab{b}})\citenamefont {Becerra}, \citenamefont {Fan},\
  and\ \citenamefont {Migdall}}]{becerra2013implementation}%
  \BibitemOpen
  \bibfield  {author} {\bibinfo {author} {\bibfnamefont {F.}~\bibnamefont
  {Becerra}}, \bibinfo {author} {\bibfnamefont {J.}~\bibnamefont {Fan}},\ and\
  \bibinfo {author} {\bibfnamefont {A.}~\bibnamefont {Migdall}},\ }\bibfield
  {title} {\bibinfo {title} {Implementation of generalized quantum measurements
  for unambiguous discrimination of multiple non-orthogonal coherent states},\
  }\href@noop {} {\bibfield  {journal} {\bibinfo  {journal} {Nature
  communications}\ }\textbf {\bibinfo {volume} {4}},\ \bibinfo {pages} {1}
  (\bibinfo {year} {2013}{\natexlab{b}})}\BibitemShut {NoStop}%
\bibitem [{\citenamefont {Ferdinand}\ \emph {et~al.}(2017)\citenamefont
  {Ferdinand}, \citenamefont {DiMario},\ and\ \citenamefont
  {Becerra}}]{ferdinand2017multi}%
  \BibitemOpen
  \bibfield  {author} {\bibinfo {author} {\bibfnamefont {A.}~\bibnamefont
  {Ferdinand}}, \bibinfo {author} {\bibfnamefont {M.}~\bibnamefont {DiMario}},\
  and\ \bibinfo {author} {\bibfnamefont {F.}~\bibnamefont {Becerra}},\
  }\bibfield  {title} {\bibinfo {title} {Multi-state discrimination below the
  quantum noise limit at the single-photon level},\ }\href@noop {} {\bibfield
  {journal} {\bibinfo  {journal} {npj Quantum Information}\ }\textbf {\bibinfo
  {volume} {3}},\ \bibinfo {pages} {1} (\bibinfo {year} {2017})}\BibitemShut
  {NoStop}%
\bibitem [{\citenamefont {Burenkov}\ \emph {et~al.}(2020)\citenamefont
  {Burenkov}, \citenamefont {Jabir}, \citenamefont {Battou},\ and\
  \citenamefont {Polyakov}}]{burenkov2020time}%
  \BibitemOpen
  \bibfield  {author} {\bibinfo {author} {\bibfnamefont {I.}~\bibnamefont
  {Burenkov}}, \bibinfo {author} {\bibfnamefont {M.}~\bibnamefont {Jabir}},
  \bibinfo {author} {\bibfnamefont {A.}~\bibnamefont {Battou}},\ and\ \bibinfo
  {author} {\bibfnamefont {S.}~\bibnamefont {Polyakov}},\ }\bibfield  {title}
  {\bibinfo {title} {Time-resolving quantum measurement enables
  energy-efficient, large-alphabet communication},\ }\href@noop {} {\bibfield
  {journal} {\bibinfo  {journal} {PRX Quantum}\ }\textbf {\bibinfo {volume}
  {1}},\ \bibinfo {pages} {010308} (\bibinfo {year} {2020})}\BibitemShut
  {NoStop}%
\bibitem [{\citenamefont {Sidhu}\ \emph {et~al.}(2021)\citenamefont {Sidhu},
  \citenamefont {Izumi}, \citenamefont {Neergaard-Nielsen}, \citenamefont
  {Lupo},\ and\ \citenamefont {Andersen}}]{sidhu2021quantum}%
  \BibitemOpen
  \bibfield  {author} {\bibinfo {author} {\bibfnamefont {J.~S.}\ \bibnamefont
  {Sidhu}}, \bibinfo {author} {\bibfnamefont {S.}~\bibnamefont {Izumi}},
  \bibinfo {author} {\bibfnamefont {J.~S.}\ \bibnamefont {Neergaard-Nielsen}},
  \bibinfo {author} {\bibfnamefont {C.}~\bibnamefont {Lupo}},\ and\ \bibinfo
  {author} {\bibfnamefont {U.~L.}\ \bibnamefont {Andersen}},\ }\bibfield
  {title} {\bibinfo {title} {Quantum receiver for phase-shift keying at the
  single-photon level},\ }\href@noop {} {\bibfield  {journal} {\bibinfo
  {journal} {PRX Quantum}\ }\textbf {\bibinfo {volume} {2}},\ \bibinfo {pages}
  {010332} (\bibinfo {year} {2021})}\BibitemShut {NoStop}%
\bibitem [{\citenamefont {Clarke}\ \emph {et~al.}(2001)\citenamefont {Clarke},
  \citenamefont {Chefles}, \citenamefont {Barnett},\ and\ \citenamefont
  {Riis}}]{clarke2001experimental}%
  \BibitemOpen
  \bibfield  {author} {\bibinfo {author} {\bibfnamefont {R.~B.}\ \bibnamefont
  {Clarke}}, \bibinfo {author} {\bibfnamefont {A.}~\bibnamefont {Chefles}},
  \bibinfo {author} {\bibfnamefont {S.~M.}\ \bibnamefont {Barnett}},\ and\
  \bibinfo {author} {\bibfnamefont {E.}~\bibnamefont {Riis}},\ }\bibfield
  {title} {\bibinfo {title} {Experimental demonstration of optimal unambiguous
  state discrimination},\ }\href@noop {} {\bibfield  {journal} {\bibinfo
  {journal} {Physical Review A}\ }\textbf {\bibinfo {volume} {63}},\ \bibinfo
  {pages} {040305} (\bibinfo {year} {2001})}\BibitemShut {NoStop}%
\bibitem [{\citenamefont {Sol{\'\i}s-Prosser}\ \emph
  {et~al.}(2017)\citenamefont {Sol{\'\i}s-Prosser}, \citenamefont {Fernandes},
  \citenamefont {Jim{\'e}nez}, \citenamefont {Delgado},\ and\ \citenamefont
  {Neves}}]{solis2017experimental}%
  \BibitemOpen
  \bibfield  {author} {\bibinfo {author} {\bibfnamefont {M.}~\bibnamefont
  {Sol{\'\i}s-Prosser}}, \bibinfo {author} {\bibfnamefont {M.}~\bibnamefont
  {Fernandes}}, \bibinfo {author} {\bibfnamefont {O.}~\bibnamefont
  {Jim{\'e}nez}}, \bibinfo {author} {\bibfnamefont {A.}~\bibnamefont
  {Delgado}},\ and\ \bibinfo {author} {\bibfnamefont {L.}~\bibnamefont
  {Neves}},\ }\bibfield  {title} {\bibinfo {title} {Experimental minimum-error
  quantum-state discrimination in high dimensions},\ }\href@noop {} {\bibfield
  {journal} {\bibinfo  {journal} {Physical review letters}\ }\textbf {\bibinfo
  {volume} {118}},\ \bibinfo {pages} {100501} (\bibinfo {year}
  {2017})}\BibitemShut {NoStop}%
\bibitem [{\citenamefont {Patterson}\ \emph {et~al.}(2021)\citenamefont
  {Patterson}, \citenamefont {Chen}, \citenamefont {Wossnig}, \citenamefont
  {Severini}, \citenamefont {Browne},\ and\ \citenamefont
  {Rungger}}]{patterson2021quantum}%
  \BibitemOpen
  \bibfield  {author} {\bibinfo {author} {\bibfnamefont {A.}~\bibnamefont
  {Patterson}}, \bibinfo {author} {\bibfnamefont {H.}~\bibnamefont {Chen}},
  \bibinfo {author} {\bibfnamefont {L.}~\bibnamefont {Wossnig}}, \bibinfo
  {author} {\bibfnamefont {S.}~\bibnamefont {Severini}}, \bibinfo {author}
  {\bibfnamefont {D.}~\bibnamefont {Browne}},\ and\ \bibinfo {author}
  {\bibfnamefont {I.}~\bibnamefont {Rungger}},\ }\bibfield  {title} {\bibinfo
  {title} {Quantum state discrimination using noisy quantum neural networks},\
  }\href@noop {} {\bibfield  {journal} {\bibinfo  {journal} {Physical Review
  Research}\ }\textbf {\bibinfo {volume} {3}},\ \bibinfo {pages} {013063}
  (\bibinfo {year} {2021})}\BibitemShut {NoStop}%
\bibitem [{\citenamefont {Dalla~Pozza}\ and\ \citenamefont
  {Caruso}(2020)}]{caruso2020qsd}%
  \BibitemOpen
  \bibfield  {author} {\bibinfo {author} {\bibfnamefont {N.}~\bibnamefont
  {Dalla~Pozza}}\ and\ \bibinfo {author} {\bibfnamefont {F.}~\bibnamefont
  {Caruso}},\ }\bibfield  {title} {\bibinfo {title} {Quantum state
  discrimination on reconfigurable noise-robust quantum networks},\ }\href
  {https://doi.org/10.1103/PhysRevResearch.2.043011} {\bibfield  {journal}
  {\bibinfo  {journal} {Phys. Rev. Research}\ }\textbf {\bibinfo {volume}
  {2}},\ \bibinfo {pages} {043011} (\bibinfo {year} {2020})}\BibitemShut
  {NoStop}%
\bibitem [{\citenamefont {Fanizza}\ \emph {et~al.}(2019)\citenamefont
  {Fanizza}, \citenamefont {Mari},\ and\ \citenamefont
  {Giovannetti}}]{fanizza2019optimal}%
  \BibitemOpen
  \bibfield  {author} {\bibinfo {author} {\bibfnamefont {M.}~\bibnamefont
  {Fanizza}}, \bibinfo {author} {\bibfnamefont {A.}~\bibnamefont {Mari}},\ and\
  \bibinfo {author} {\bibfnamefont {V.}~\bibnamefont {Giovannetti}},\
  }\bibfield  {title} {\bibinfo {title} {Optimal universal learning machines
  for quantum state discrimination},\ }\href@noop {} {\bibfield  {journal}
  {\bibinfo  {journal} {IEEE Transactions on Information Theory}\ }\textbf
  {\bibinfo {volume} {65}},\ \bibinfo {pages} {5931} (\bibinfo {year}
  {2019})}\BibitemShut {NoStop}%
\bibitem [{\citenamefont {Kurzy{\'n}ski}\ and\ \citenamefont
  {W{\'o}jcik}(2013)}]{kurzynski2013quantum}%
  \BibitemOpen
  \bibfield  {author} {\bibinfo {author} {\bibfnamefont {P.}~\bibnamefont
  {Kurzy{\'n}ski}}\ and\ \bibinfo {author} {\bibfnamefont {A.}~\bibnamefont
  {W{\'o}jcik}},\ }\bibfield  {title} {\bibinfo {title} {Quantum walk as a
  generalized measuring device},\ }\href@noop {} {\bibfield  {journal}
  {\bibinfo  {journal} {Physical review letters}\ }\textbf {\bibinfo {volume}
  {110}},\ \bibinfo {pages} {200404} (\bibinfo {year} {2013})}\BibitemShut
  {NoStop}%
\bibitem [{\citenamefont {Li}\ \emph {et~al.}(2019)\citenamefont {Li},
  \citenamefont {Zhang},\ and\ \citenamefont {Zhu}}]{li2019implementation}%
  \BibitemOpen
  \bibfield  {author} {\bibinfo {author} {\bibfnamefont {Z.}~\bibnamefont
  {Li}}, \bibinfo {author} {\bibfnamefont {H.}~\bibnamefont {Zhang}},\ and\
  \bibinfo {author} {\bibfnamefont {H.}~\bibnamefont {Zhu}},\ }\bibfield
  {title} {\bibinfo {title} {Implementation of generalized measurements on a
  qudit via quantum walks},\ }\href@noop {} {\bibfield  {journal} {\bibinfo
  {journal} {Physical Review A}\ }\textbf {\bibinfo {volume} {99}},\ \bibinfo
  {pages} {062342} (\bibinfo {year} {2019})}\BibitemShut {NoStop}%
\bibitem [{\citenamefont {Whitfield}\ \emph {et~al.}(2010)\citenamefont
  {Whitfield}, \citenamefont {Rodr{\'\i}guez-Rosario},\ and\ \citenamefont
  {Aspuru-Guzik}}]{whitfield2010quantum}%
  \BibitemOpen
  \bibfield  {author} {\bibinfo {author} {\bibfnamefont {J.~D.}\ \bibnamefont
  {Whitfield}}, \bibinfo {author} {\bibfnamefont {C.~A.}\ \bibnamefont
  {Rodr{\'\i}guez-Rosario}},\ and\ \bibinfo {author} {\bibfnamefont
  {A.}~\bibnamefont {Aspuru-Guzik}},\ }\bibfield  {title} {\bibinfo {title}
  {Quantum stochastic walks: A generalization of classical random walks and
  quantum walks},\ }\href@noop {} {\bibfield  {journal} {\bibinfo  {journal}
  {Physical Review A}\ }\textbf {\bibinfo {volume} {81}},\ \bibinfo {pages}
  {022323} (\bibinfo {year} {2010})}\BibitemShut {NoStop}%
\bibitem [{\citenamefont {Fujiwara}\ \emph {et~al.}(2003)\citenamefont
  {Fujiwara}, \citenamefont {Takeoka}, \citenamefont {Mizuno},\ and\
  \citenamefont {Sasaki}}]{fujiwara2003exceeding}%
  \BibitemOpen
  \bibfield  {author} {\bibinfo {author} {\bibfnamefont {M.}~\bibnamefont
  {Fujiwara}}, \bibinfo {author} {\bibfnamefont {M.}~\bibnamefont {Takeoka}},
  \bibinfo {author} {\bibfnamefont {J.}~\bibnamefont {Mizuno}},\ and\ \bibinfo
  {author} {\bibfnamefont {M.}~\bibnamefont {Sasaki}},\ }\bibfield  {title}
  {\bibinfo {title} {Exceeding the classical capacity limit in a quantum
  optical channel},\ }\href@noop {} {\bibfield  {journal} {\bibinfo  {journal}
  {Physical review letters}\ }\textbf {\bibinfo {volume} {90}},\ \bibinfo
  {pages} {167906} (\bibinfo {year} {2003})}\BibitemShut {NoStop}%
\bibitem [{\citenamefont {Cerf}\ \emph {et~al.}(2002)\citenamefont {Cerf},
  \citenamefont {Bourennane}, \citenamefont {Karlsson},\ and\ \citenamefont
  {Gisin}}]{cerf2002security}%
  \BibitemOpen
  \bibfield  {author} {\bibinfo {author} {\bibfnamefont {N.~J.}\ \bibnamefont
  {Cerf}}, \bibinfo {author} {\bibfnamefont {M.}~\bibnamefont {Bourennane}},
  \bibinfo {author} {\bibfnamefont {A.}~\bibnamefont {Karlsson}},\ and\
  \bibinfo {author} {\bibfnamefont {N.}~\bibnamefont {Gisin}},\ }\bibfield
  {title} {\bibinfo {title} {Security of quantum key distribution using d-level
  systems},\ }\href@noop {} {\bibfield  {journal} {\bibinfo  {journal}
  {Physical review letters}\ }\textbf {\bibinfo {volume} {88}},\ \bibinfo
  {pages} {127902} (\bibinfo {year} {2002})}\BibitemShut {NoStop}%
\bibitem [{\citenamefont {Degen}\ \emph {et~al.}(2017)\citenamefont {Degen},
  \citenamefont {Reinhard},\ and\ \citenamefont {Cappellaro}}]{degen2017}%
  \BibitemOpen
  \bibfield  {author} {\bibinfo {author} {\bibfnamefont {C.}~\bibnamefont
  {Degen}}, \bibinfo {author} {\bibfnamefont {F.}~\bibnamefont {Reinhard}},\
  and\ \bibinfo {author} {\bibfnamefont {P.}~\bibnamefont {Cappellaro}},\
  }\bibfield  {title} {\bibinfo {title} {Quantum sensing},\ }\href@noop {}
  {\bibfield  {journal} {\bibinfo  {journal} {Rev. Mod. Phys.}\ }\textbf
  {\bibinfo {volume} {89}},\ \bibinfo {pages} {035002} (\bibinfo {year}
  {2017})}\BibitemShut {NoStop}%
\bibitem [{\citenamefont {Lambert}\ \emph {et~al.}(2013)\citenamefont
  {Lambert}, \citenamefont {Chen}, \citenamefont {Cheng}, \citenamefont {Li},
  \citenamefont {Chen},\ and\ \citenamefont {Nori}}]{lambert2013quantum}%
  \BibitemOpen
  \bibfield  {author} {\bibinfo {author} {\bibfnamefont {N.}~\bibnamefont
  {Lambert}}, \bibinfo {author} {\bibfnamefont {Y.-N.}\ \bibnamefont {Chen}},
  \bibinfo {author} {\bibfnamefont {Y.-C.}\ \bibnamefont {Cheng}}, \bibinfo
  {author} {\bibfnamefont {C.-M.}\ \bibnamefont {Li}}, \bibinfo {author}
  {\bibfnamefont {G.-Y.}\ \bibnamefont {Chen}},\ and\ \bibinfo {author}
  {\bibfnamefont {F.}~\bibnamefont {Nori}},\ }\bibfield  {title} {\bibinfo
  {title} {Quantum biology},\ }\href@noop {} {\bibfield  {journal} {\bibinfo
  {journal} {Nature Physics}\ }\textbf {\bibinfo {volume} {9}},\ \bibinfo
  {pages} {10} (\bibinfo {year} {2013})}\BibitemShut {NoStop}%
\bibitem [{\citenamefont {Schuld}\ \emph {et~al.}(2014)\citenamefont {Schuld},
  \citenamefont {Sinayskiy},\ and\ \citenamefont
  {Petruccione}}]{schuld2014quest}%
  \BibitemOpen
  \bibfield  {author} {\bibinfo {author} {\bibfnamefont {M.}~\bibnamefont
  {Schuld}}, \bibinfo {author} {\bibfnamefont {I.}~\bibnamefont {Sinayskiy}},\
  and\ \bibinfo {author} {\bibfnamefont {F.}~\bibnamefont {Petruccione}},\
  }\bibfield  {title} {\bibinfo {title} {The quest for a quantum neural
  network},\ }\href@noop {} {\bibfield  {journal} {\bibinfo  {journal} {Quantum
  Information Processing}\ }\textbf {\bibinfo {volume} {13}},\ \bibinfo {pages}
  {2567} (\bibinfo {year} {2014})}\BibitemShut {NoStop}%
\end{thebibliography}%


\newpage
\clearpage

\section{Supplemental material}

\section{Theoretical Framework}
The state of the network can be represented as a 4-dimensional vector, where the first two dimensions represents the input layer and the second pair represents the intermediate layer, i.e. the sinker nodes.
The sink nodes are considered as an auxiliary system, subjected to measurement and not evolving in time.
Since in our model a superposition state between the input layer and the sinker nodes is not allowed, the only possible states have the form $\begin{pmatrix}
\alpha\\
\beta\\
0\\
0
\end{pmatrix}$ or$\begin{pmatrix}
0\\
0\\
\gamma\\
\delta
\end{pmatrix}$.
Therefore, the one-step evolution of the system is described by a 4x4 unitary matrix, composed of two antidiagonal blocks $U=\begin{pmatrix}
0 & 0 & a & b\\
0 & 0 & c & d\\
a' & b' & 0 & 0\\
c' & d' & 0 & 0
\end{pmatrix}$. Indeed, also the permanence of the system in the same layer after an evolution step is not allowed by our model. 
The left-bottom block, which we refer to as $U_F$, represents the forward evolution of the system from input layer nodes $\{1,2\}$ towards the sinker nodes $\{3,4\}$, while the right-top block, which we refer to as $U_B$, represents the backward evolution of the system from sinker nodes $\{3,4\}$ to the input layer nodes $\{1,2\}$.
In the actual implementation, both layers are encoded in the polarization degree of freedom, while they have a different temporal and spatial localization. Therefore, in the main text we usually refer to the evolution as an alternate application of operators $\hat{U}_F$ and $\hat{U}_B$ to the same 2-D vector.
These two matrices can be identically or differently set, producing a symmetric or asymmetric network.
In our framework, we are also able to discriminate the extraction time of the photon, i.e. the amount of times the photon has completed a loop evolution before being sent to the sinks.
This supplemental degree of freedom, which is crucial in our protocol, can be formally addressed as two auxiliary systems, one for the photon travelling through the network, representing the evolution step of the system, and one for the sinks, representing the extraction time.
In conclusion, the evolution of the composite system of the network system $n$ and the sink system $s$ , besides the time degree of freedom, can be segmented in three different phases, given the initial state $ \ket{\psi_0}_{n,s}=(\alpha_0\ket{1}_n+\beta_0\ket{2}_n)\ket{t_0}_n $. In the first one, the initial state evolves from input to intermediate layer through $\hat{U}_F$:
\begin{equation*}
    \ket{\psi_0^{(1)}}_{n,s}=(\hat{U}_F\otimes \hat{I}) \ket{\psi_0}_{n,s}= (\gamma_0\ket{3}_n+\delta_0\ket{4}_n)\ket{t_0}_n, 
\end{equation*}
The second step consists of the extraction to the sink nodes, which corresponds to the generation of a superposition state of the system being in the network or in the sinks at a certain step time, through a projector from the network to the sinks:
\begin{equation*}
\begin{split}
    \ket{\psi_0^{(2)}}_{n,s}=\bigg[\sqrt{T}\big(\hat{I}_n\otimes \hat{I}_s\big)+\\
    +\sqrt{1-T}\big(\ket{5}_s\bra{3}_n+\ket{6}_s\bra{4}_n\big)\otimes\bigg(\sum_{k=0}^\infty \ket{t_k}_s\bra{t_k}_n\bigg) \bigg]\ket{\psi_0^{(1)}}=\\ =\sqrt{T}\big(\gamma_0\ket{3}_n+\delta_0\ket{4}_n\big)\ket{t_0}_n+\sqrt{1-T}\big(\gamma_0\ket{5}_s+\delta_0\ket{6}_s\big)\ket{t_0}_s,
\end{split}    
\end{equation*}

where $T$ is the probability of the system to stay in the network and $1-T$ the probability of being extracted to the sinks.
The last step brings again the network in a state of the input layer, while the sinks do not evolve at all:
\begin{equation*}
\begin{split}
    \ket{\psi_1}_{n,s}=\bigg(\hat{U}_B\otimes\sum_{k=0}^\infty \ket{t_{k+1}}_n\bra{t_k}_n\bigg) \ket{\psi_0^{(2)}}_{n,s}=\\ =\sqrt{T}\big(\alpha_1\ket{1}_n+\beta_1\ket{2}_n\big)\ket{t_1}_n+\sqrt{1-T}\big(\gamma_0\ket{5}_s+\delta_0\ket{6}_s\big)\ket{t_0}_s, 
\end{split}    
\end{equation*}
where the "time state" of the network has been updated, because a forward and backward evolution has been completed.
The evolution continues as a repetition of these three steps, leading to a general state, after $M$ completed loops:
\begin{equation*}
\begin{split}
    \ket{\psi_M}=\sqrt{T}^M\big(\alpha_M\ket{1}_n+\beta_M\ket{2}_n\big)\ket{t_M}_n+\\
    +\sum_{k=1}^M\sqrt{T}^{k-1} \sqrt{1-T}\big(\gamma_{k-1}\ket{5}_s+\delta_{k-1}\ket{6}_s\big)\ket{t_{k-1}}_s,
\end{split}    
\end{equation*}
where $\alpha_k,\beta_k,\gamma_k, \delta_k$ with $k=0,..,M$ are the coefficients for each basis state of the network after a $k$ steps evolution.
The measurement we are able to perform, in the end, corresponds to a projection on both the sink state and the corresponding step time, which determines the probability of finding the system in a given sink $\sigma$ at a certain time $t_k$:
\begin{equation*}
    P_\sigma\big(t_k\big)=\abs{\bra{\sigma}_s\bra{t_k}_s\ket{\phi}_s}^2
\end{equation*}
where $\ket{\phi}_s$ is a general joint state of the sinks and the associated extraction time.
The probability distributions computed in this way correspond to the ones reported in the main text, which are crucial to our analyses. This formalism allows us to intuitively describe the capability of our setup to exploit time as a further degree of freedom for discrimination, while the actual realization is implemented through a set of subsequent post-selection procedures, which are described in detail in the next sections.

\section{Optimization Methods}
In the case of binary quantum state discrimination, the one-shot strategy consisting of projecting the two states to the orthogonal basis states produces the optimal performance, achieving the Helstrom bound.
In the same framework, the natural generalization of this method to a multi-step strategy is a protocol performing a Helstrom bound level discrimination at any step, namely an optimal information extraction each time the system occurs in the sinker nodes (the intermediate layer).
The most simple realization of such a protocol consists of generating an "optimal" output state at the first extraction step, through the tuning of operator $\hat{U}_F$, and then tailoring $\hat{U}_B$ to produce the same state at any extraction step.
In conclusion, $\hat{U}_F$ must be the optimally discriminating projector and $\hat{U}_B$ must be such that $\hat{U}_F\circ\hat{U}_B=\hat{I}$.
In our case, $U_F=U_B=\frac{1}{\sqrt{2}}\begin{pmatrix}
1 & 1\\
1 & -1
\end{pmatrix}$ with results reported in the main text.
In this case, the network layers are connected by a symmetrical transformation; the exploitation of a globally asymmetrical network unitary, with $U_F\neq U_B$, allows for more advanced applications, such as four state discrimination.
In particular, this method is effective for a set of geometrically homogeneous states, since the network parameters are fixed in time.
In our case of study, where the set to be discriminated is $\{\ket{+},\ket{-},\ket{R},\ket{L}\}$, the network unitary is tailored with the aim of producing the most different output probability distributions as the four different states occur at the input layer.
That has been made selecting one of the two sinks and requesting that in each of the first four time bins a different input state had the relative maximum probability of being extracted. 
The evolution matrices produced following this method are $U_F=\frac{1}{\sqrt{2}}\begin{pmatrix}
1 & 1\\
1 & -1
\end{pmatrix}$ and $U_B=\frac{1}{\sqrt{2}}\begin{pmatrix}
\frac{1+i}{2} & \frac{1+i}{2}\\
\frac{1-i}{2} & \frac{-1+i}{2}
\end{pmatrix}$.
The product matrix, representing the loop evolution of the system between two distinct extraction steps, results then $U_L=U_F\circ U_B=\frac{1}{\sqrt{2}}\begin{pmatrix}
1 & i\\
i & 1
\end{pmatrix}$.
It is worth noting that $U_L^4=I$, from which the 4-periodicity of the output probability distributions derives. Therefore, the first four time bins contain all of the information on the input state, while the experimental exploitation of the further ones is only useful for collecting a greater amount of meaningful signal.
Since the states are geometrically homogeneous, separated by a $\frac{\pi}{2}$ rotation around the same axis, this kind of procedure produces an optimal output.
In fact, without changing the network parameters in time, it is possible to achieve a $P_{err}=0.5$, which is the analytical bound for a set of four geometrically homogeneous states \cite{yuen1975optimum}.
 \begin{figure}[!h]
        \includegraphics[width=\columnwidth]{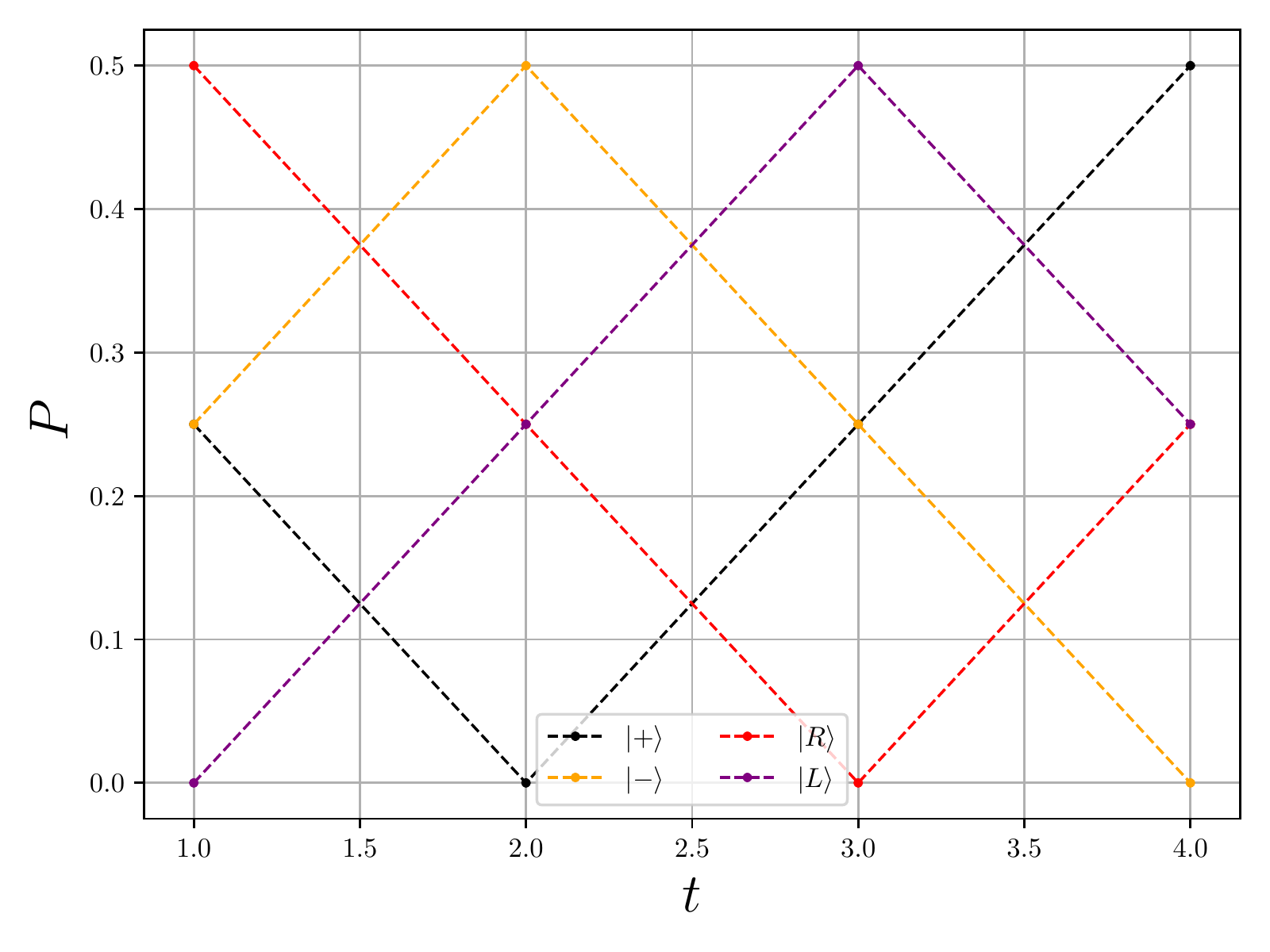}\\
        \includegraphics[width=\columnwidth]{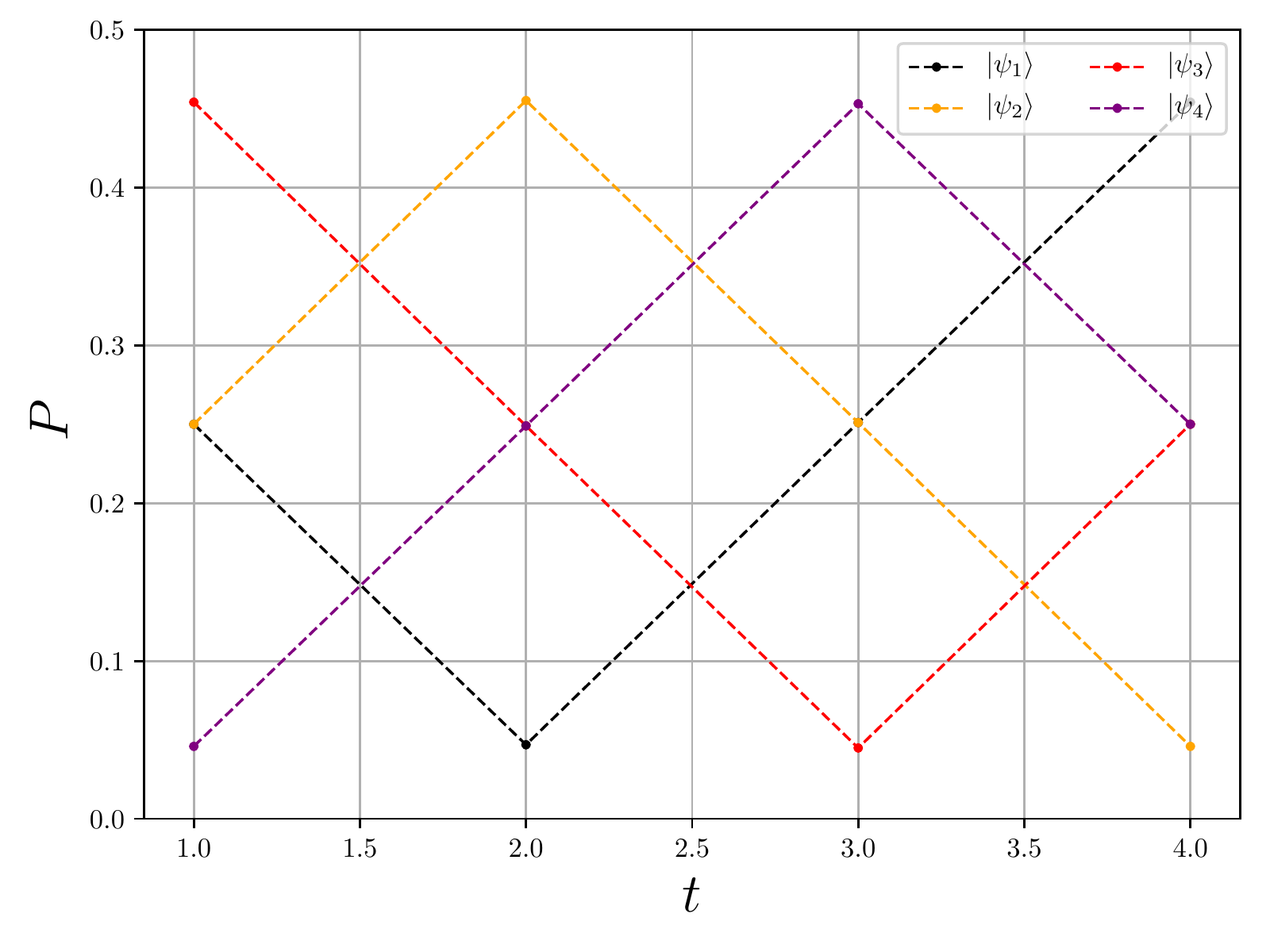}
        \caption{\textbf{Output probability distributions for both states sets.} \textit{Time-wise output probability of the first four extraction steps when the geometriccaly homogeneous states (top) and the Tetrad (bottom) states circulate in the network.
        } }
        \label{fig:time_sig}
\end{figure}
This is understandable by looking at Fig.\ref{fig:time_sig} (top), where the numerical probability of detection in the first four time bins is displayed for the geometrically homogeneous set and the normalization is performed without considering the total signal decrease which the system experiences after every extraction step. Moreover, the probability distributions are normalized in such a way that, given a certain input state, the total probability equals to $1$. In this way, the error probability of guess can be straightforwardly computed, given the input state. 
The results of the same analysis are shown in Fig. \ref{fig:time_sig} (bottom) for the Tetrad set.
For this latter set, the network optimization was carried according to the same method, aiming at maximizing the discrimination probability between different pairs of states at each extraction step. 
The corresponding evolution matrices, resulting from a numerical maximization,  were
\begin{align*}
U_F=\frac{1}{\sqrt{2}}\begin{pmatrix}
0.953021  & 0.302905 \\
-0.302905  & 0.953021
\end{pmatrix}\\
U_B=\frac{1}{\sqrt{2}}\begin{pmatrix}
-0.674645 + 0.216571 i & 0.2118 - 0.673121 i\\
-0.2118 - 0.673121i & -0.674645 - 0.216571 i
\end{pmatrix}
\end{align*}
which, again, produce a loop evolution $U_L\sim\frac{1}{\sqrt{2}}\begin{pmatrix}
1 & i\\
i & 1
\end{pmatrix}$
In this case, the method produces a $P_{err}>0.5$, since there are not orthogonal subsets in the Tetrad set, but still a very significant result. In fact, as the number of available copies increases, the error probability scales down exponentially, as discussed below.

\section{Experimental setup details}
Photon pairs are generated by a high brilliance SPDC source realized according to the model described in \cite{fedrizzi2007wavelength}: a PPKTP crystal, embedded in a Sagnac interferometer, pumped by a single mode CW laser radiation ($\lambda_p = 405$nm), which generates collinear pair of photons (\textit{system} and \textit{ancilla}) with opposite polarization at a wavelength $\lambda_{i,s} = 2\lambda_p = 810$nm . They are both coupled to a pair of optical fibers, but headed to different outputs: the \textit{system} photon is injected into the setup and actually undergoes the network evolution, while the \textit{ancilla} photon is directly sent to a single-photon detector, acting as a trigger for coincidences.
After collection of both photons, their detection is processed by ID-Quantique time tagger ID800: this device features the possibility of setting a narrow coincidences window (up to 81 ps), the capability of recording the relative detection time of a photon, hence the possibility of electronically setting any delay between two photon counters with the aim of computing coincidences and also to perform high-resolved time scanning of the delay between two detecting channels.
Thanks to these features, we were able to implement the time-binning discrimination strategy, which is theoretically fancied above.
It is worth mentioning that the first considered extraction step is the second, because of the different extraction probability featured by the first one: in fact, an unbalanced Beamsplitter (BS) was exploited for the extraction step, featuring a transmittivity $T\sim70\%$ and a reflectivity $R\sim30\%$. Therefore, as understandable from Fig.2 of the main text, the first extraction step features $T$ as extraction probability, while the remaining ones are characterized by $R$. Because of that, we chose to neglect the coincidences of photons extracted at the first step.

\section{Data analysis}

The experimental verification of the effectiveness of the time-binning strategy for multi-state discrimination consisted of the experimental reconstruction of the expected output probability distributions. In our framework, the reconstruction consisted of recording the amount of photons being extracted in one of the two sinks after any amount of travelled loops.
Since the source generates photon pairs in a non-deterministic fashion, the only way to be sure the counted photons had travelled the right distance in the setup consisted of exploiting a coincidence measurement.
Thanks to the ID800 by ID-Quantique, it is possible to count coincidences at any delay between the system and ancilla photon. Therefore, it was possible to observe the amount of photons extracted at each step by suitably tuning the considered delay, depending on the path difference between photons generated within the same temporal window.
The measured coincidences had then to be cleaned off the background noise: as described above, the first actual detection step, not taken into account in the analysis, features a much higher signal with respect to the subsequent ones. Therefore, it causes a high amount of accidental coincidences for any set delay. A measurement of the background noise due to the first extraction photons is performed for each delay and directly subtracted to the time-binned measurements, in order to get time-wise coincidences profiles which only display proper coincidences. %A supplemental cleaning is performed by subtracting the coincidences detected for very long delays, corresponding to a travelled distance in the setup such that the actual system photons had completely been extracted or absorbed. Therefore, these coincidences could only be due to noise.\\
In conclusion, clean step-wise coincidences counts were obtained, for each input state of both the considered sets. Because of the high amount of detected coincidences, the resulting output was considered as an average result per se. Therefore, the output probability distributions for each input states, displayed in the main text, were directly deduced by the normalization of the clean coincidences profiles.

\section{Low Photon Number}
\begin{figure}[!h]
        \includegraphics[width=\columnwidth]{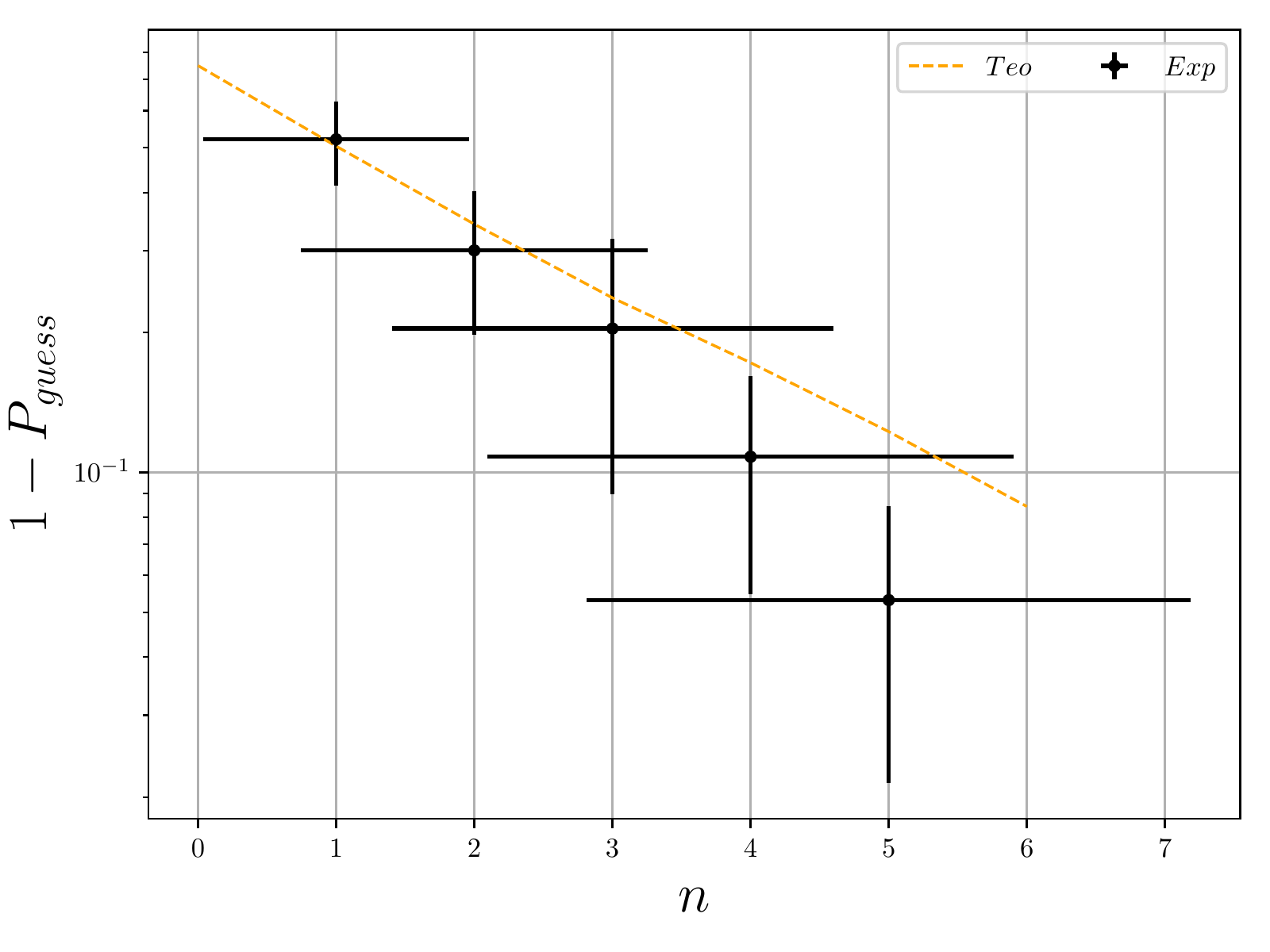}\\
        \includegraphics[width=\columnwidth]{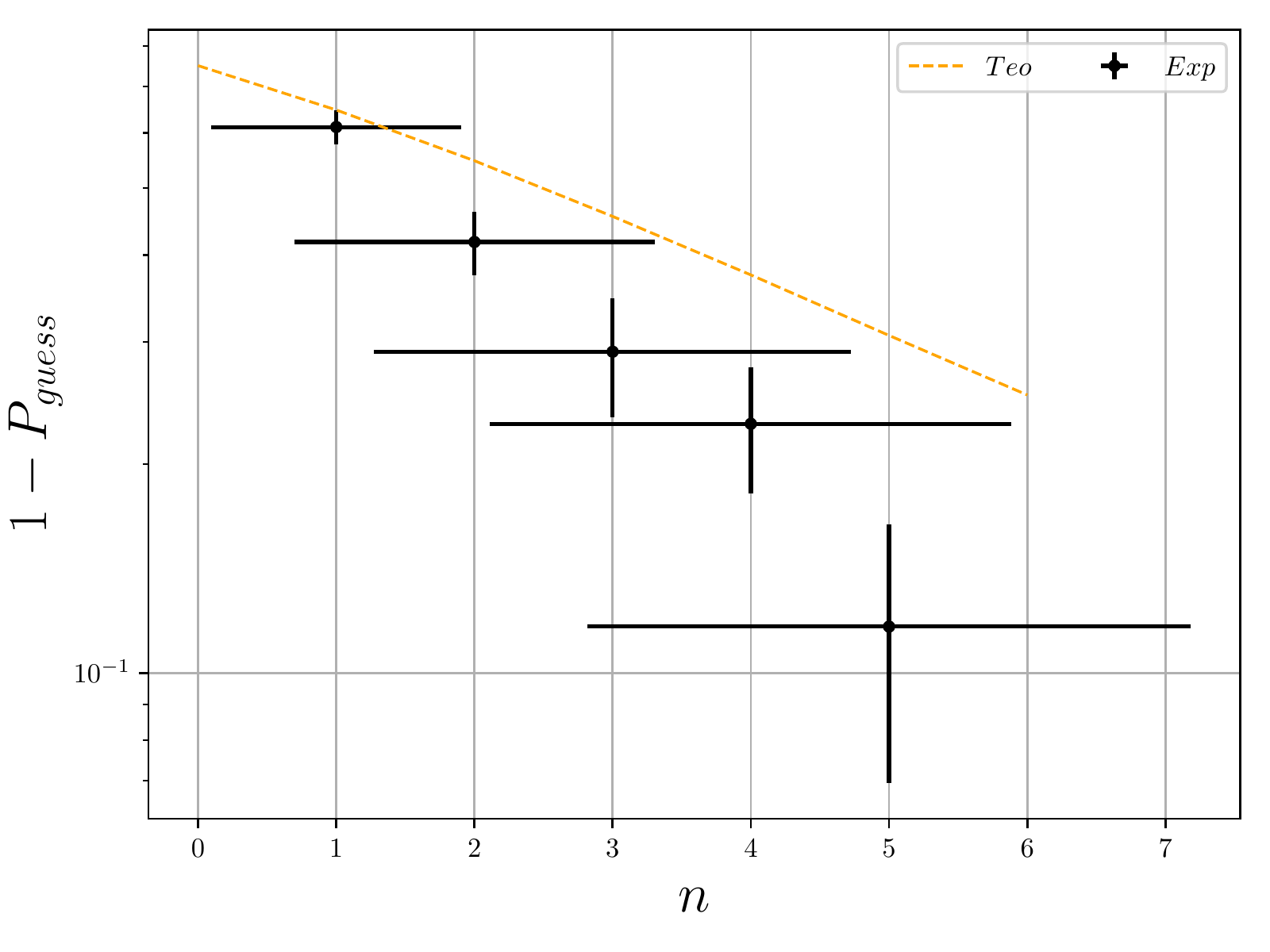}
        \caption{\textbf{Error probability scaling for both states sets.} \textit{Scaling of the error probability in function of the number of analyzed copies of the system, for the geometrically homogeneous states (top) and the Tetrad  states (bottom). It is worth noting that, in these specific cases, the experimental asymmetries produce an enhanced scaling of the probability of correct detection as the number of exploited photons grow.
        } }
        \label{fig:scaling}
\end{figure}

In order to verify the effectiveness of the multi-state discrimination protocol in an actual scenario, we tuned the photon source depicted above to a low average photon number regime: through this source, it is not possible to deterministically generate each photon, but rather it is possible to set an average time rate of single photon generation. 
Nevertheless, through this procedure, it is possible to test the average quality of the protocol by computing the average error probability $P_{err}$ for a single copy of the system and how this quantity scales as the average available number of copies grows.
The experimental strategy to compute the average $P_{err}$ started from the setting of the average rate of photon per second generated by the source. An average rate of $1$-$2$ total observed coincidences per second was set, and a higher average photon number was obtained by a longer time integration.
In this regime, the same time-binned measurements exploited to evaluate the step-wise output distributions can be performed, registering $n$-seconds events for different integration time intervals $n$. We address as \textit{$n$-seconds event} a time-binned measurement performed for $n$ seconds.
The events measured in this way have still to be cleaned off the accidental coincidences, as mentioned in the previous section. Hence, it is not possible to consider single instances of $n$-seconds events, but rather an average $n$-seconds event has to be taken into account, to get the chance of subtracting background noise (which is only meaningful as an average quantity).
Since the SPDC photon generation is subjected to thermal noise, a $n$-seconds measurement does not always corresponds to an event featuring a number of detected coincidences predictable by mere Poissonian statistics.
Therefore, a post-selection procedure on the number of registered coincidences has to be performed in order get actually comparable events. In this way, we only take into account for our analysis those events that reasonably feature the same generation conditions, turning $n$-seconds events into $k$-photons events. 
In conclusion, a set of $n$-seconds events was observed, they were post-selected to feature a photon count compatible with a Poissionian statistics and then they were averaged obtaining an average $k$-photons event. A corresponding average noise event was subtracted to each $k$-photon event, producing a set of average clean $m$-photon events (in general, subtracting the noise reduces the total number of detected coincidences). Given an input state $\psi_i$ from the set $\{\psi_j\}_{j=1}^4$ and an average clean $m$-photon event $\bar{E}_m$, the probability of error in guessing the input state was computed as:
\begin{equation*}
    P_{err}(\psi_i, m)=1-P(\psi_i|\bar{E}_m)=1-\frac{P(\bar{E}_m|\psi)}{\sum_{\psi_j}P(\bar{E}_m|\psi_j)}
\end{equation*}
exploiting the Bayes's rule.
The resulting trends of $P_{err}$ in function of the number of detected photons are shown in Fig. \ref{fig:scaling}, for both sets of states.

\end{document}